\documentclass[11pt]{article}
\usepackage[left=2cm, right=2cm, top=2cm, bottom=2cm]{geometry}

\usepackage{graphicx} 
\graphicspath{{./}}
\usepackage{subcaption}
\usepackage[round]{natbib}
\usepackage{amsmath}
\usepackage{hyperref}
\usepackage{orcidlink}
\usepackage{amsfonts}
\usepackage{booktabs}
\usepackage{algorithm}
\usepackage{algorithmic}
\usepackage{mathtools}
\mathtoolsset{showonlyrefs} 
\usepackage{amsthm}

\DeclarePairedDelimiter\floor{\lfloor}{\rfloor}
\DeclarePairedDelimiter\ceil{\lceil}{\rceil}

\DeclareMathOperator{\N}{\mathbb{N}}
\DeclareMathOperator{\E}{\mathbb{E}}
\DeclareMathOperator{\SI}{\mathbb{S}}
\DeclareMathOperator{\R}{\mathbb{R}}
\newcommand{\allocation}{\kappa}
\newcommand{\midprice}{p}
\newcommand{\inventory}{Q}
\newcommand{\activeorders}{m}
\newcommand{\traderstd}{\delta}
\newcommand{\bI}{\bar{I}}
\newcommand{\action}{a}

\theoremstyle{remark}

\title{Multi-Level Market Making with Reinforcement Learning\thanks{Support from Swiss NSF Grant 10003723 is gratefully acknowledged.}}
\author{Patrick Cheridito\,\orcidlink{0000-0001-9074-7295} and
	Moritz Weiss\,\orcidlink{0009-0004-6024-1313}\thanks{Corresponding author: moritz.tilmann.weiss@gmail.com} \\
	Department of Mathematics, ETH Zurich, Switzerland \\[1.5em]
	\textit{Keywords}: Trading, market making, limit order book, reinforcement learning
	\\[1.5em]
	\textit{JEL Classification}: C45, C61, C63, G17
	}

\date{}

\begin{document}

\maketitle 

\begin{abstract}
We introduce a reinforcement learning framework for market making in a limit order book. Our algorithm aims to maximize trading revenue by dynamically submitting market and limit orders of varying sizes across multiple price levels while controlling inventory size. We use multivariate logistic-normal distributions to model order allocations and employ a deep-set encoder to aggregate features from variable-length order sets into a fixed-dimensional latent representation. Additionally, we incorporate potential-based reward shaping to accelerate learning without altering the optimal policy. We illustrate the performance of the method in three simulated market environments consisting of noise traders who submit random trades, tactical traders who respond to instantaneous volume imbalance, and strategic traders who trade in the direction of an exponentially weighted volume 
imbalance signal.
\end{abstract}

\section{Introduction}

Market makers operate by placing orders in a limit order book to maximize expected revenue. 
Unlike directional traders, who accumulate exposure to price trends, market makers capture the 
bid-ask spread while actively trying to keep their inventory 
small to reduce their market risk. The market-making problem has been studied extensively in the mathematical 
finance literature, but typically under stylized models of asset price dynamics and fill probabilities 
as well as simplified action spaces to keep the problem analytically tractable. E.g.\
\citet{ho1981optimal, avellaneda2008high, gueant2013dealing, guilbaud2013optimal, cartea2020market} model the 
mid-price together with fill probabilities and consider a market maker who only uses one
limit buy and sell order; see also \cite{cartea2015algorithmic} for a review of the market-making and trade-execution literature.

The market-making problem has also been tackled with reinforcement learning (RL). But most approaches to date have worked with simplified action spaces. In \cite{gavsperov2021market}, the market maker attempts to find the optimal distance from the mid-price of a single limit buy and sell order. Similarly, in \cite{guo2023market}, the market maker chooses the distance of limit orders from a reservation price and places a single unit on each side of the book. In \cite{shi2024market}, the market maker chooses the distance of limit orders from the mid-price, places unit-size limit orders and learns a separate hedging rate. In \cite{beysolow2019market, sun2022market, chung2022market, kumar2023deep, zheng2024reinforcement}, the market maker chooses discrete actions, which correspond to different order placements. Some works use more complex action spaces. E.g.\ in \cite{abergel2022algorithmic}, the market maker controls the prices of a fixed number of limit orders across multiple order book levels, but their optimization method is computationally expensive. The work \cite{wang2024market} employs a Beta distribution to parameterize the allocation of limit order volume across price levels, but the market maker posts the same volume on both sides of the book. 

One of the main challenges in applying RL to trading is designing realistic simulation environments. Backtesting with historical data requires researchers to artificially inject order fills, and this
still fails to capture feedback loops caused by other market participants' responses to the agent's submitted 
limit orders. To overcome this limitation, recent literature has leveraged stochastic market simulators that 
organically interact with the RL agent and provide a controlled training as well as testing environment. While several works have adopted this paradigm to train market-making RL agents \citep{xiong2015comparison, lu2018order, gavsperov2022deep, lalor2025event}, our approach uses a multi-agent simulation based on Poisson order-flow modeling \citep{cont2010stochastic,cheridito2026reinforcement}.

The main contributions of our paper are: 
\begin{itemize}
\item The formulation of the market-making problem with a general state space and action space. The state space incorporates the full limit order book, queue positions, as well as past order flows. The action space allows the market maker to post market and limit orders of varying sizes across multiple price levels. 

\item The development of an actor-critic RL algorithm for market making. The algorithm uses a multivariate logistic-normal distribution enabling flexible placement, a deep-set encoder to handle variable-length features, and potential-based reward shaping to accelerate learning. 

\item Detailed numerical experiments evaluating the performance of the algorithm in three simulated market environments populated by noise, tactical, and strategic trading agents, exhibiting adverse selection behavior.
\end{itemize}

\section{Limit order books}
\label{sec:limit_order_books}

Most modern exchanges use {\sl limit order books} (LOBs) to match buyers and sellers. They record all 
outstanding limit orders, which represent commitments to buy or sell a given quantity at a stated price.
Quantities are quoted in multiples of a {\sl lot}, the smallest tradable unit, and
prices move on a discrete grid whose increment is called a {\sl tick}. Buy orders (bids) queue on one side of 
the book and sell orders (asks) on the other.  The {\sl best quotes} at time $t$ are given by the 
{\sl highest bid} $p^b_t$ and {\sl lowest ask} $p^a_t$. They are separated by the {\sl spread} 
$p^a_t - p^b_t > 0$. The mid-price $p_t =(p^b_t +p^a_t)/2$ is usually used as the reference price 
of the asset. A trade occurs when someone submits a market order, which is immediately matched 
with the best limit orders on the other side of the book. Submitted limit orders are 
not guaranteed to be executed and can also be cancelled. The dynamics of the LOB are described by
\begin{equation}
	v^{b}_t = (v^{b,1}_t, \dots, v^{b,D}_t) \in \N^D_0
	\quad \text{and} \quad 
	v^{a}_t = (v^{a,1}_t, \dots, v^{a,D}_t) \in \N^D_0,
\label{eq:volume} 
\end{equation} 
where $D\in\N$, $v^{b,k}_t$ is the number of lots $k-1$ ticks below the best bid price
and $v^{a,k}_t$ the number of lots $k-1$ ticks above the best ask price. By definition, the volumes 
$v^{b,1}_t$ and $v^{a,1}_t$ at the best quotes are always positive, but the other volumes could also be zero.
Orders at each price level form queues and are executed on a first-come, first-served basis. The price level 
$l$ of a limit order is its distance from the best price in ticks, using a one-tick index shift (so $l = 1$ corresponds to the best price). The queue position $q$ at a given price level is the number of lots with higher priority on the same queue, also indexed with a one-unit shift (so $q = 1$ denotes the highest-priority order, which will be matched first if a market order reaches that price level). 
A limit sell (buy) order at a given price level is filled by a market buy (sell) order if the market 
order's volume is large enough to consume all limit orders at more aggressive price levels and those 
ahead in the queue at that price level. Partial fills occur when the remaining market order volume is 
insufficient to fully fill the limit order; a partially filled order stays in the book with its volume reduced accordingly. Both price level and queue position determine a limit order's fill probability and are 
therefore critical for market-making strategies. In principle, there is no fixed price limit for 
submitting limit orders, but in our simulation, we observe only the first $D$ levels on the bid and ask sides of the book. The volume vector \eqref{eq:volume} reflects supply and demand in the market and can indicate short-term price pressure. For example, more volume on the bid than ask side indicates buying pressure resulting in an increased likelihood that the next price movement is upward. For a more detailed discussion of limit order books, 
we refer to the review paper by \citet{gould2013limit}.

\section{The market-making problem}
\label{sec:market_making_problem}

We train a market-making algorithm to trade during a given time interval $[0,T].$ 
The algorithm acts at the discrete times $t_n = n \Delta t$ for $n = 0,1,\dots,N-1,$
where $N\in\N$ and $\Delta t = T/N.$ At each time $t_n,$ the algorithm observes a market 
state $s_n,$ takes an action $\action_n$ and receives a reward $r_n(s_n, \action_n).$ In the following, we drop the subscript $n$ from $r_n$ and write $r(s_n, \action_n)$ since in our setting, $t_n$ is contained in the state $s_n$
together with other information available at $t_n$. Actions $\action_n$ are taken immediately after $t_n$
and consist of cancellations of old limit orders together with placements of new market and limit orders. 
We constrain the algorithm to allocate at most $M\in\N$ lots to market and limit orders at each decision time $t_n.$ The reward $r(s_n, \action_n)$ comprises the cash flows from market and limit order fills during
$(t_{n}, t_{n+1}]$. The algorithm starts with an inventory $\inventory_0=0$.
$\inventory_n$ is the difference of the total number of the algorithm's buy and sell order fills in the interval $[0,t_n]$ measured in lots. At terminal time $t_N=T$, the algorithm receives a terminal reward $g(s_N)$ depending on the terminal state $s_N$. The inventory $\inventory$ can change by at most $M$ lots from time $t_n$ to $t_{n+1}$. For example, if the algorithm sends $M$ market buy orders, the inventory increases by $M$. 
The algorithm's goal is to learn a stochastic policy, modeled as a conditional density $\pi$ on the action space, 
that maximizes the expected reward
\begin{equation} 
	J(\pi) = \underset{\substack{\action\sim\pi \\ s_0\sim\rho}}{\E} \left[ \sum_{n={0}}^{N-1} r(s_n, \action_n) + g(s_N) \right]	.
    \label{eq:optimization_objective} 
\end{equation} 
The subscript $s_0\sim \rho$ indicates that the initial state $s_0$ is sampled from a distribution $\rho$, while
$\action\sim\pi$ means that each action $\action_n$ in the sequence $\action=(\action_0, \action_1, \dots, \action_{N-1})$ is sampled from a conditional distribution with density $\pi(\cdot\mid s_n)$. \cite{avellaneda2008high} have studied both finite- and infinite-horizon formulations of the market-making problem. Here, we focus on the finite-horizon version \eqref{eq:optimization_objective}.

\subsection{State space}
\label{sec:state_space} 
At each decision time $t_n$, $n = 0,1,\dots,N-1$, the algorithm observes market states that
are visible to all market participants and private states only known to the algorithm.
 
\paragraph{Market states}
The market states consist of the following quantities: 
\begin{itemize}
\item 
the best bid and ask prices $p^b_n$ and $p^a_n$,
\item 
the first $K \in\N$ entries of the two volume vectors defined in \eqref{eq:volume},
\item 
the mid-price drift over $(t_{n-1}, t_n]$ given by $\Delta p_n = p_n - p_{n-1}$,
\item 
the market order flow $\Delta^M_n$ defined as the difference between the total volume of market 
buy and sell orders in the interval $(t_{n-1}, t_n]$,
\item 
the limit order flow $\Delta^L_n$ defined as the difference between the total volume of limit buy 
and sell orders placed during the interval $(t_{n-1}, t_n]$,
\item 
the cancellation order flow $\Delta^C_n$ defined as the difference between the total 
volume of limit buy and sell orders cancelled during the interval $(t_{n-1}, t_n]$.
\end{itemize}

\paragraph{Private states} 
The private states contain the quantities:

\begin{itemize} 
\item the current decision time $t_n$ during the algorithm's trading interval $[0,T]$,
\item the algorithm's inventory $\inventory_n$,
\item the numbers $\activeorders^b_n$ and $\activeorders^a_n$ of the algorithm's remaining buy 
and sell limit orders at time $t_n$,
\item the levels, queue positions and sizes 
    \begin{equation}
    \label{eq:queue_position}
    (l^{b,i}_n, q^{b,i}_n, w^{b,i}_n) \in \N \times \N \times \N, \; i = 1, \dots, m^b_n,
    \quad 
    (l^{a,j}_n, q^{a,j}_n, w^{a,j}_n) \in \N \times \N \times \N, \; j = 1, \dots, m^a_n,
    \end{equation}
    of the algorithm's remaining limit orders at time $t_n$, where
    $(l^{b,i}_n, q^{b,i}_n, w^{b,i}_n) =(l,q,w)$ means that the $i$-th limit buy order consists of 
    $w$ lots, is resting $l-1$ ticks below the best bid, and there are $q-1$ other lots on the same price level with higher priority. Similarly, $(l^{a,j}_n, q^{a,j}_n, w^{a,j}_n) =(l,q,w)$ means that the $j$-th limit sell order consists of $w$ lots, is resting $l-1$ ticks above the best ask, and there are $q-1$ other lots on the same price level with higher priority. 
\item the algorithm's remaining limit orders at time $t_n$ expressed as fractions of $M$
\begin{equation}
\allocation_n = \left(
\allocation^{b,1}_n, \dots, \allocation^{b,K+1}_n,
\allocation^{a,1}_n, \dots, \allocation^{a,K+1}_n \right)\in [0, 1]^{2(K+1)}.
\label{eq:order_per_price_level}
\end{equation}
Here, $\allocation^{b,k}_n$ (or $\allocation^{a,k}_n$), $k = 1, \dots, K$, are the fractions of $M$ 
lots sitting $k-1$ ticks below the best bid (or above the best ask) and $\allocation^{b,K+1}_n$ (or $\allocation^{a,K+1}_n$) is the fraction of $M$ lots at least $K$ ticks below (or above) the best bid (or ask) price. 
\end{itemize}

We normalize the features above to speed up the learning, as explained in Appendix~\ref{sec:feature_normalization}. In principle, the feature \eqref{eq:order_per_price_level} is redundant, as the information is already contained in \eqref{eq:queue_position}, but we add it since we must compute it to allocate orders, as explained in the following section, and including this feature helps with the training of the algorithm.

\subsection{Action space}
\label{sec:action_space}
The algorithm's actions $\action_n$ live in the simplex 
\begin{equation}
\SI^{2(K+1)}=
\left\{\action=(\action^0, \action^1, \dots, \action^{2(K+1)}) \, : \,
\sum_{k=0}^{2(K+1)} \action^{k} = 1 \, , \,
\action^{k}\geq 0 \right\}.
\label{eq:simplex}
\end{equation}
The components of the vector $\action_n$ describe the market and limit order allocation of the algorithm
right after time $t_n$. Remaining limit orders from $(t_{n-1}, t_n]$ are cancelled and reallocated 
only when necessary. When reallocating, the algorithm first cancels orders with lower priority.
$\action^{0}$ denotes the fraction of $M$ lots that is not placed in the market, while
$\action^{1}$ is the fraction allocated to market buy orders. For $k = 2, \dots, K+1$, $\action^{k}$ denotes the fraction of $M$ lots allocated to buy limit orders placed $k-2$ ticks below the best bid price. 
$\action^{K+2}$ is the fraction allocated to market sell orders. For $k = K+3, \dots, 2(K+1)$, $\action^{k}$ denotes the fraction allocated to sell limit orders placed $k-K-3$ ticks above the best ask price. 

The quantities $\action^{k} M$ have to be rounded to integer values. We do this with the \textsl{Hamilton apportionment method} \citep{balinski2010fair}. That is, we first reduce $\action^{k} M$ to their integer 
parts $\floor*{\action^{k} M}$. This typically results in a total allocated volume less than $M$. The remaining volume is then distributed sequentially to the component with the largest fractional remainder, then to the component with the second-largest remainder and so on until the total allocated volume equals $M$. 

\begin{figure}[htbp]
    \centering
    \includegraphics[scale=0.9]{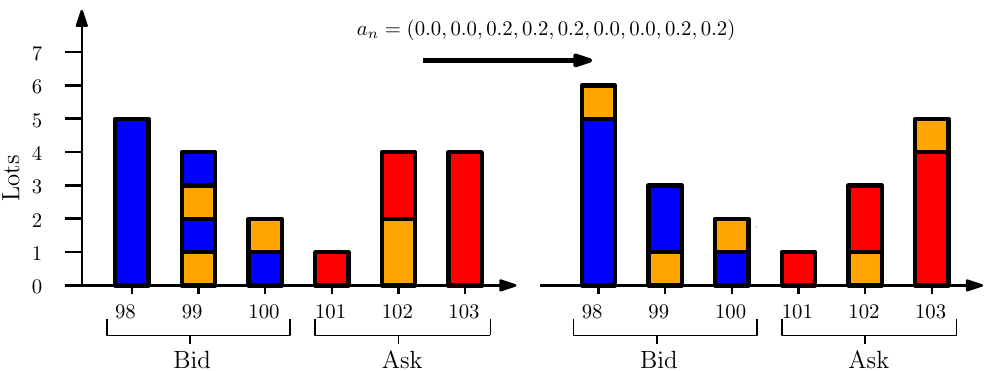}
    \caption{
    Order book before and after the action $\action_n$ has been taken. The blue boxes indicate limit buy orders, the red boxes indicate limit sell orders, and orange boxes represent the algorithm's orders. }
    \label{fig:action_and_state}
\end{figure}

In Figure~\ref{fig:action_and_state}, we illustrate how the algorithm's actions act on the order book. 
For simplicity, we assume the algorithm has $M=5$ lots available to allocate as market or limit orders
at the first $K=3$ levels on both sides of the book. Suppose that at time $t_n$, the order book volumes are 
\begin{equation} 
(v^{b,1}_n, v^{b,2}_n, v^{b,3}_n) = (2, 4, 5)
\quad \text{and} \quad
(v^{a,1}_n, v^{a,2}_n, v^{a,3}_n) = (1,4,4),
\end{equation}
and the algorithm's own limit buy and sell orders are given by 
\begin{align}
(l^{b,1}_n, q^{b,1}_n, w^{b,1}_n) &=(1,2,1), \; 
(l^{b,2}_n, q^{b,2}_n, w^{b,2}_n) =(2,1,1), \;
(l^{b,3}_n, q^{b,3}_n, w^{b,3}_n) =(2,3,1), \\ 
(l^{a,1}_n, q^{a,1}_n, w^{a,1}_n) &= (2,1,2).
\end{align}
Then, the algorithm takes the action
\[ 
\action_n =(0.0, 0.0, 0.2, 0.2, 0.2, 0.0, 0.0, 0.2, 0.2) \in \SI^8,
\] 
that is, it sends no market orders and reallocates the limit orders so that
it has one lot on each of the first three levels of the bid side and levels 2 and 3 of the ask side.
So, right after time $t_n$, the volumes in the book are
\begin{equation} 
(v^{b,1}_{n+}, v^{b,2}_{n+}, v^{b,3}_{n+})= (2, 3, 6)
\quad \text{and} \quad 
(v^{a,1}_{n+}, v^{a,2}_{n+}, v^{a,3}_{n+}) =(1,3,5),
\end{equation} 
and the algorithm's orders are given by
\begin{align}
(l^{b,1}_{n+}, q^{b,1}_{n+}, w^{b,1}_{n+}) &= (1,2,1), \;
(l^{b,2}_{n+}, q^{b,2}_{n+}, w^{b,2}_{n+}) = (2,1,1), \;
(l^{b,3}_{n+}, q^{b,3}_{n+}, w^{b,3}_{n+}) = (3,6,1), \\ 
(l^{a,1}_{n+}, q^{a,1}_{n+}, w^{a,1}_{n+}) &= (2,1,1), \;
(l^{a,2}_{n+}, q^{a,2}_{n+}, w^{a,2}_{n+}) = (3,5,1).
\end{align}

\subsection{Rewards}
\label{sec:rewards}

Let $\bar{r}(s_n,\action_n)$ be the cash flows from the market and limit order fills in the interval $(t_n, t_{n+1}],$ and write $\midprice_n$ and $\inventory_n$ for the mid-price and inventory at time $t_n$. Then, we define the reward function by
\begin{equation}
\label{eq:rewards}
r(s_n, \action_n) = \frac{1}{M}\left( \bar r(s_n, \action_n) + (\inventory_{n+1}p_{n+1} - \inventory_n p_n) - \gamma \left\lvert \inventory_{n+1} \right\rvert \right)\quad \text{for} \; n=0,1,\dots,N-1,
\end{equation}
with an inventory risk parameter $\gamma\geq 0.$ At terminal time $t_N,$ we impose the position limit $|\inventory_{N+}|\leq\ceil{\nu M}$ for $\nu \in[0,1],$ where $\ceil{\nu M}$ denotes the smallest integer greater than or equal to $\nu M.$ Any inventory beyond this limit is liquidated by a market order. More precisely, the algorithm sends a market sell order or market buy order of size $\max(|\inventory_N| - \ceil{\nu M},0)$, depending on whether $\inventory_N$ is positive or negative. The cash flow from this market order is denoted by $\text{MO}_\nu(s_N)$, which depends on the full terminal state $s_N$ because the market order walks through the limit order volumes resting in the book at $t_N$. Let us denote the left-over inventory at time $t_N+$ right after sending the market order by $\inventory_{N+}$, which is given by 
\begin{equation}
	\inventory_{N+} = \text{sign}(\inventory_N)\min\left(|\inventory_N|, \ceil{\nu M}\right).
\end{equation}
We value the left-over inventory at the mid-price and define the terminal reward by
\begin{equation}
g(s_N) = \frac{1}{M} \big(
p_N(\inventory_{N+} - \inventory_N) + \text{MO}_\nu(s_N) \big). 
\label{eq:terminal_reward}
\end{equation}
Then, we have
\begin{equation}
\label{eq:telescoping}
\sum_{n=0}^{N-1} r(s_n, \action_n) + g(s_N) = \frac{1}{M}\left(\sum_{n=0}^{N-1} \big( \bar r(s_n, \action_n) - \gamma \lvert \inventory_{n+1} \rvert \big)
+ \inventory_{N+} p_N
+ \text{MO}_\nu(s_N)\right).
\end{equation}

The expression $\inventory_{n+1}p_{n+1}-\inventory_n p_n$ evaluates the change of the algorithm's wealth from time $t_n$ to time $t_{n+1},$ where the algorithm's inventory is valued at the mid-price. Although this term cancels when rewards are summed over the episode \eqref{eq:telescoping}, we include it in the reward because it helps the algorithm to learn the optimal policy faster. This is a form of potential-based reward shaping \citep{ng1999potentialreward}, which preserves optimal policies while providing denser learning signals.
If we were to use only the cash flows together with the inventory penalty as a reward, the price movements affecting held inventory would contribute to the reward only at terminal time $t_N$, leading to delayed and sparse feedback and making credit assignment difficult. The expression $-\gamma |\inventory_{n+1}|$ penalizes inventory holdings, which is common in the market-making literature \citep{cartea2015algorithmic}. Including such an inventory penalty leads to a lower standard deviation of profits, at the cost of lower expected profits, as discussed in Appendix~\ref{sec:effect_of_inventory_penalty}.

\section{Actor-critic policy gradient algorithm} 
\label{sec:actor-critic-algorithm}

We use an actor-critic algorithm to find an optimal stochastic policy, called {\sl actor}, by estimating the gradient of the objective function $J(\cdot)$, defined in \eqref{eq:optimization_objective}, with respect to the policy parameters. A value function estimate, called {\sl critic}, is used to reduce the variance of the gradient estimate. In our algorithm, we also use an encoder that can handle variable-length queue position features. Therefore, our actor-critic algorithm consists of three components: the encoder, the actor, and the critic. Each component will be explained in more detail in this section. 

\subsection{Encoder}

The order features \eqref{eq:queue_position} represent the level, queue position, and size of the algorithm's resting limit orders. The number of resting limit orders varies over time, but neural networks need fixed-dimensional inputs. One way to obtain a fixed-dimensional representation is to use a vector of triples (level, queue position, and size) of length $2M$, where the entries $1, \dots, M$ correspond to the $m^b_n$ active limit buy orders and the entries $M+1, \dots, 2M$ to the $m^a_n$ active limit sell orders. Any entries beyond $m^b_n$ and $m^a_n$ are padded with zeros. 

To illustrate this representation, we consider an example. Suppose that the algorithm has $M=5$ lots available and places them on up to $K=3$ price levels. Furthermore, we suppose that the buy and sell order features are given by  
\[
[(1,1,1), (3,1,1), (0,0,0), (0,0,0), (0,0,0)] \quad \text{and} \quad 
[(1,1,3), (0,0,0), (0,0,0), (0,0,0), (0,0,0)].
\]
Now we suppose that the algorithm decides to allocate two lots to the best bid and one lot to the second-best bid, by cancelling the sell order resting on the best ask, while keeping the remaining order allocations the same. Furthermore, we suppose that the order queues on the best and second-best bid each contain three lots. After reallocation, the buy and sell order features are given by
\[
[(1,1,1), (1,4,2), (2,4,1), (3,1,1), (0,0,0)] 
\quad \text{and} \quad 
[(0,0,0), (0,0,0), (0,0,0), (0,0,0), (0,0,0)].
\]
Importantly, this changes the ordering of the buy orders. The order with level, queue position, and size $(3,1,1)$ was originally in the second slot of the buy order features, but after reallocation it is now in the fourth slot. The orders with level, queue position, and size $(1,4,2)$ and $(2,4,1)$ were not present in the original order features, but they are now in the second and third slots. For a standard feed-forward network, each parameter at the first layer is associated with a specific input coordinate. Consequently, when an unchanged order moves from one coordinate to another, the network encounters the same information at a different input coordinate.

To avoid this issue, we use a deep-set encoder to represent limit orders. Deep-set encoders \citep{zaheer2017deep} operate on set-valued inputs by applying a shared embedding function to each element and aggregating the resulting embeddings using a permutation-invariant operation such as summation or averaging.
Let $f_\phi^o$ be a neural network with parameters $\phi$. For $k\in\{1,\dots,K\}$, we define the index sets
\[
I_n^{b,k} = \big\{i\in\{1,\dots,m_n^b\}:l_n^{b,i}=k\big\},
\qquad
I_n^{a,k} = \big\{i\in\{1,\dots,m_n^a\}:l_n^{a,i}=k\big\}.
\]
Then, we define the encoded order features on the buy and sell side at price level $k$ by
\begin{equation} 
f_{n, \phi}^{e,b,k} =
\frac{1}{|I_n^{b,k}|}
\sum_{i\in I_n^{b,k}}
f_\phi^o\Big(q_n^{b,i},w_n^{b,i}\Big)
\quad 
\text{and} 
\quad
f_{n, \phi}^{e,a,k} =
\frac{1}{|I_n^{a,k}|}
\sum_{i\in I_n^{a,k}}
f_\phi^o\Big(q_n^{a,i},w_n^{a,i}\Big).
\label{eq:queue_position_encoding}
\end{equation}
If $I_n^{b,k}$ or $I_n^{a,k}$ is empty, the corresponding encoded feature is set equal to zero. Since averaging is invariant to permutations, the representation does not depend on the way in which individual limit orders are stored. At the same time, queue priority is retained because the queue position $q$ enters the order-level network $f_\phi^o$. By averaging, we lose information about the total volume of orders at each price level, but this is already captured by the feature \eqref{eq:order_per_price_level}. We denote by $f^e_\phi(s_n)$ the encoded state vector, which contains the encoded order features $f_{n, \phi}^{e,b,k}$ and $f_{n, \phi}^{e,a,k}$ for $k = 1,\dots,K $ together with the remaining features from Section~\ref{sec:state_space}. 		

Continuing with our example, the encoded buy and sell features before and after reallocation are given by 
\[
[f_{n, \phi}^{e,b,1}, f_{n, \phi}^{e,b,2},  f_{n, \phi}^{e,b,3}] 
\quad \text{and} \quad 
[f_{n, \phi}^{e,a,1}, f_{n, \phi}^{e,a,2},  f_{n, \phi}^{e,a,3}]. 
\]
Before reallocation, the first and third entries of the buy features are non-zero, while the second entry is zero, and the first entry of the sell features is non-zero, while the last two entries are zero. After reallocation, all three entries of the buy features are non-zero, and all three entries of the sell features are zero.
Crucially, the interpretation of the entries does not change. The first entry always describes orders at the first price level, the second entry always describes orders at the second price level, and the third entry always describes orders at the third price level, on the buy and sell side. The encoder therefore converts a variable-size collection of orders into a fixed-dimensional representation, where the $k$-th slot on the buy or sell side always corresponds to order information on the $k$-th price level.

\subsection{Actor} 
In our application, the actions must be contained in the simplex. A natural choice for the distribution parameterizing the actions is the {\sl logistic-normal distribution}, which has support in $\SI^{2(K+1)}$. The logistic transformation $h$ maps vectors $x=(x^1, x^2, \dots, x^{2(K+1)})$ in $\R^{2(K+1)}$ to actions $\action=(\action^0, \action^1, \dots, \action^{2(K+1)})$ in $\SI^{2(K+1)},$ where
\begin{align}
	\action^k &= \frac{e^{x^k}}{1 + \sum_{l=1}^{2(K+1)} e^{x^l} }, \quad \text{for} \quad k = 1,\dots,2(K+1), \\
	\action^{0} &= \frac{1}{1+\sum_{l=1}^{2(K+1)} e^{x^l}}.  
\end{align}
 Let $X$ be a random variable with multivariate normal distribution on $\R^{2(K+1)}$ with mean vector $\mu=\left(\mu^1, \dots, \mu^{2(K+1)}\right) \in \R^{2(K+1)}$ and covariance matrix $\Sigma = (\Sigma^{ij})_{i,j = 1,\dots,2(K+1)} \in \R^{2(K+1)\times 2(K+1)}.$ Then, the random variable $h(X)$ has a logistic-normal distribution, which admits a closed-form density \citep{atchison1980logistic}.
\newcommand{\tv}{{\theta^{v}}}
\newcommand{\tm}{{\theta^{m}}}
We choose a policy (or actor) that is given by a logistic-normal distribution with conditional density $\pi_{\phi,\theta}(\cdot\mid s)$ for each state $s$, where $\phi$ are the parameters of the encoder network and $\theta=(\tm, \tv)$ are the weights specifying the mean and variance of the distribution. More precisely, $\tm$ are the weights of a neural network $f^m_\tm$ that maps encoded states $f^e_\phi(s)$ to the mean of the normal distribution $\mu_\tm = f^m_\tm(f^e_\phi(s)),$ and we model the covariance matrix as a state-independent diagonal matrix given by 
\begin{equation}
\Sigma_{\tv}=\mathrm{Diag}\left(\exp\left(\theta^{v,1}\right),\exp\left(\theta^{v,2}\right), \dots, \exp\left(\theta^{v,2(K+1)}\right)\right) \in \R^{2(K+1)\times 2(K+1)} 
.
\label{eq:covariance}
\end{equation}

The mean of the logistic-normal distribution is not known in closed form. However, if an action $\action$ has logistic-normal distribution with density $\pi_{\phi,\theta}(\cdot\mid s)$, where $\mu_\tm$ is the mean of the underlying normal distribution, we have
\begin{align}
	\E \left[ \log \left( \frac{\action^j}{\action^k} \right) \right] &= \mu_\tm^j - \mu_\tm^k, 
    \quad
    \text{for} \; j, k \in \{0, 1, 2, \dots,2(K+1)\},
    \label{eq:mean}
\end{align}
with the convention $\mu^0_\tm=0.$ At initialization, we set the parameters of the final layer of the network $f^m_{\tm}$ close to zero, except for the bias term $b\in\R^{2(K+1)}.$ Then, the bias $b$ controls the policy's initial allocations via \eqref{eq:mean}.

\subsection{Critic}
The value function (or critic) $V^{\pi_{\phi,\theta}}$ and the advantage function $A^{\pi_{\phi,\theta}}$ corresponding to the actor $\pi_{\phi,\theta}$ are defined by
\begin{align}
        \label{eq:value_function}
        V^{\pi_{\phi,\theta}}(s^\prime) &= \underset{\action\sim\pi_{\phi,\theta}}{\E} \left[ \sum_{l=n}^{N-1} r(s_l, \action_l) + g(s_N) \mid s_n =s^\prime \right], \\
		A^{\pi_{\phi,\theta}}(s^\prime, \action^\prime) &=
		\underset{\action\sim\pi_{\phi,\theta}}{\E} \left[ \sum_{l=n}^{N-1} r(s_l, \action_l) + g(s_N) \mid s_n =s^\prime, \action_n=\action^\prime \right]-
        V^{\pi_{\phi,\theta}}(s^\prime),
        \label{eq:advantage} 
\end{align}
where $s^\prime$ is a state and $\action^\prime$ an action at time step $n.$ The advantage function is estimated based on an estimate of the value function. Let $f^V_\vartheta$ be a neural network with weights $\vartheta$. Then, the value function estimate for a state $s^\prime$ is given by $V^{\pi_{\phi,\theta}}_{\phi,\vartheta}(s^\prime)= f^V_{\vartheta}(f^e_\phi(s^\prime)).$ For a given episode
$\{(s_n,\action_n) : n \in \{0,1,\dots,N-1 \} \} \cup \{s_N\},$ we estimate the advantage function by
\begin{equation}
    A_{\phi,\vartheta}^{\pi_{\phi,\theta}}(s_n, \action_n) =
    \sum_{l=n}^{N-1} r(s_l, \action_l) + g(s_N) - V^{\pi_{\phi,\theta}}_{\phi, \vartheta}(s_n), \quad \text{for} \quad n = 0,1,\dots,N-1,
    \label{eq:advantage_estimate}
\end{equation}
replacing the expected value in \eqref{eq:advantage} by the returns-to-go in this episode and the value function with its estimate. 

\subsection{Training process} 
\label{sec:training_process}

By the policy gradient theorem \citep{sutton1999policy}, the gradient of the objective \eqref{eq:optimization_objective} with respect to the parameters of the policy is given by
\begin{align}
		\nabla_{\phi, \theta} J(\pi_{\phi, \theta}) =
		\underset{\substack{\action \sim \pi_{\phi,\theta} \\ s_0 \sim \rho}}{\E}
		\left[\sum_{n=0}^{N-1} A^{\pi_{\phi,\theta}}(s_n, \action_n) \nabla_{\phi,\theta}
		\log \left(\pi_{\phi,\theta}(\action_n \mid s_n ) \right)
		\right].
\label{eq:policy_gradient}
\end{align}
In practice, we replace the true advantage $A^{\pi_{\phi,\theta}}$ by the estimator $A_{\phi,\vartheta}^{\pi_{\phi,\theta}}$ from \eqref{eq:advantage_estimate}. Since the density of the logistic-normal distribution is known in closed form, the gradient with respect to the parameters of the logistic-normal distribution can be computed efficiently \citep{cheridito2026reinforcement}. The expected value in \eqref{eq:policy_gradient} is not known in closed form, and we estimate it by sampling trajectories from the policy. The encoder and policy parameters $(\phi, \theta)$ are updated using gradient descent and an empirical estimate of the gradient expression \eqref{eq:policy_gradient}. Additionally, the parameters $\vartheta$ of the critic are updated in each gradient step.

We make $H\in\N$ gradient steps with respect to the encoder, actor, and critic parameters $\phi, \theta,\vartheta.$ At the beginning of the training cycle, the parameters are initialized as $\phi_1, \theta_1,\vartheta_1.$ In each training step $i\in\{1,\dots,H\},$ we collect $\tau\in\N$ trajectories 
\begin{equation}
	\mathcal{T}_{N, \tau}=\bigg\{
		\big\{			
		(s_{n,k}, \action_{n,k}, r(s_{n,k}, \action_{n,k})) : n\in\{0, \dots, N-1 \} 
		\big\} 
		\cup 
		\big\{(s_{N, k}, g(s_{N,k})) \big\} 
		:
		k\in\{1,\dots,\tau\}
		\bigg\},
    \label{eq:batch_of_trajectories}
\end{equation}
where the actions $\action_{n,k}$ are sampled from a logistic-normal distribution which has conditional density $\pi_{\phi_i, \theta_i}(\cdot \mid s_{n,k}).$ For each state-action pair $(s_{n,k},\action_{n,k}),$ the advantage function $A_{\phi_i, \vartheta_i}^{\pi_{\phi_i, \theta_i}}(s_{n,k}, \action_{n,k})$ is estimated via \eqref{eq:advantage_estimate} using the value function estimate $V_{\phi_i, \vartheta_i}^{\pi_{\phi_i, \theta_i}}(s_{n,k})$.
The encoder, actor, and critic parameters are updated from $\phi_i, \theta_i, \vartheta_i$ to $\phi_{i+1}, \theta_{i+1}, \vartheta_{i+1}$
by making a gradient step with learning rate $\eta>0$ using the combined loss function 
\begin{align}
(\phi, \theta, \vartheta)
\mapsto &- 
\frac{1}{\tau N}
\sum_{k=1}^\tau 	 
\sum_{n=0}^{N-1}
A_{\phi_i, \vartheta_i}^{\pi_{\phi_i, \theta_i}}(s_{n,k}, \action_{n,k})
\log \pi_{\phi, \theta}(\action_{n,k} \mid s_{n,k} ) \\ 
&+ 
c_V
\frac{1}{\tau N}
\sum_{k=1}^\tau
\sum_{n=0}^{N-1} \left\lvert V^{\pi_{\phi_i, \theta_i}}_{\phi, \vartheta}(s_{n,k}) - \sum_{l=n}^{N-1} r(s_{l,k}, \action_{l,k}) - g(s_{N,k}) \right\rvert^2,
\label{eq:policy_gradient_loss} 
\end{align}
for a parameter $c_V\in\R_+.$ Alternatively, the parameters can be updated in two separate gradient steps: one for the encoder and actor parameters $(\phi, \theta)$ and one for the encoder and critic parameters $(\phi, \vartheta) $. However, we found that the combined loss function \eqref{eq:policy_gradient_loss} yields good results, while being computationally more efficient. Algorithm~\ref{alg:actor_critic} summarizes the whole training process. 
\begin{algorithm}[htbp]
	\caption{Actor-critic algorithm with logistic-normal distribution and order encoding.}
	\begin{algorithmic}
		\STATE 
        {\textbf{Initialize} encoder weights $\phi$, actor weights $\theta$, critic weights $\vartheta,$ bias $b,$ number of iterations $H,$ number of trajectories $\tau,$ learning rate $\eta,$ inventory parameter $\gamma,$ loss function parameter $c_V.$}
		\FOR{$i = 1, \dots, H$}
		\STATE \textbf{Collect} trajectories $\mathcal{T}_{N,\tau} $ 
		by sampling actions from policy $\pi_{\phi_i,\theta_i}$. 
		\STATE{\textbf{Update} weights $\phi, \theta, \vartheta$ with a gradient step using learning rate $\eta$, the loss function \eqref{eq:policy_gradient_loss}, and the samples $\mathcal{T}_{N,\tau}.$} 
		\ENDFOR
	\end{algorithmic}
\label{alg:actor_critic}
\end{algorithm}

\section{Market simulation}
\label{sec:market_simulation_details} 

We test our algorithm in a market environment populated by three types of traders: noise traders who place and cancel orders randomly, tactical traders who react to volume imbalance, and strategic traders who trade in the direction of a smoothed volume imbalance signal. Tactical traders cause abrupt price jumps, while strategic traders cause sustained price drifts. Both effects make the market-making problem harder. All traders submit orders in the interval $[-\Delta t, T].$ The market simulation is similar to the one in \cite{cheridito2026reinforcement}. Trading agents whose order arrivals follow Poisson processes have been used extensively in the literature (see, e.g.,~\citealp{cont2010stochastic}).

\subsection{Noise traders}
\label{sec:noise_trader}
The noise traders submit market, limit, and cancellation orders according to independent Poisson processes. The intensities of market and limit orders are state-independent, and cancellation intensities scale linearly with the volume at each price level. 
\begin{itemize}
	\item Market buy and sell orders arrive with intensity $\lambda^M.$ 
	\item 
	For $k \in \{1, \dots, D\},$ limit buy and sell orders arrive $k$ ticks below the best ask price or $k$ ticks above the best bid price with intensity $\lambda^{L,k}.$ 
	\item If the spread is $j$ ticks, then for $ k \in \{j, \dots, D\},$ cancellations of limit buy orders arrive $k$ ticks below the best ask price with intensity $\lambda^{C,k} v^{b,{k-j+1}}$, and cancellations of limit sell orders arrive $k$ ticks above the best bid price with intensity $\lambda^{C,k} v^{a,{k-j+1}}.$ 	
\end{itemize}

\subsection{Tactical traders}
The intensities of the tactical traders' order arrivals depend on volume imbalance. We define exponentially weighted volumes on the bid and ask sides by
\begin{equation}
	V^b_t = \sum_{k=1}^{D} v^{b,k}_t e^{-c(k-1)} \quad \text{and} \quad V^a_t = \sum_{k=1}^{D} v^{a,k}_t e^{-c(k-1)}, 
    \label{eq:weighted_volumes}
\end{equation}
where $c \in \R_+$ is a damping factor. The weighted volume imbalance is then defined by 
\begin{equation} 
	I_t = \frac{V^b_t-V^a_t}{V^b_t+V^a_t}.
    \label{eq:volume_imbalance}
\end{equation}
The damping factor $c$ controls how sensitive the imbalance is to volumes posted at deeper levels in the book. Let $I^+_t$ and $I^-_t$ be the positive and negative parts of the imbalance. The intensities for different order types are defined as follows:
\begin{itemize}
	\item For $d^M \in \R_+,$ market buy orders arrive with intensity $d^MI^+_t$ and market sell orders arrive with intensity $d^MI^-_t.$ 
	\item For $k\in\{1,2, \dots, D\}$ and $ d^{L,k} \in \R_+$, 
    limit buy orders arrive $k$ ticks below the best ask price with intensity $d^{L,k}I^+_t,$ and limit sell orders arrive $k$ ticks above the best bid price with intensity $d^{L,k}I^-_t.$ 
	\item If the spread is $j$ ticks, then for $k\in\{j,\dots,D\}$ and $d^{C,k}\in\R_+$, cancellations of limit buy orders arrive $k$ ticks below the best ask price with intensity $d^{C,k}I^-_t v^{b,{k-j+1}}$, and cancellations of limit sell orders arrive
    $k$ ticks above the best bid price with intensity $d^{C,k}I^+_t v^{a,{k-j+1}}.$ 
\end{itemize}

\subsection{Strategic traders}
\label{sec:strategic_trader}
The strategic traders trade in the direction of a smoothed volume imbalance signal. More precisely, let the volume imbalance $I_t$ be defined as in \eqref{eq:volume_imbalance}, with the same damping factor $c.$ Then, the exponentially weighted volume imbalance signal is defined by
\begin{equation}
    \bI_t = e^{-\beta(t+\Delta t)}\, I_{-\Delta t} + \beta \int_{-\Delta t}^{t} e^{-\beta(t-s)}\, I_s\, ds, \quad t\geq-\Delta t,
	\label{eq:smoothed_imbalance}
  \end{equation}
with a parameter $\beta\in\R_+$ that controls the degree of smoothing in the signal. With a high value for $\beta,$ the signal $\bI_t$ tracks $I_t$ closely, and with a low value for $\beta,$ the signal $\bI_t$ is a heavily smoothed version of the past values of $I_t.$ The strategic traders' order arrivals then have the following intensities:

\begin{itemize}
	\item For $z^M \in \R_+,$ market buy orders arrive with intensity $z^M\bI^+_t$ and market sell orders arrive with intensity $z^M\bI^-_t.$ 
	\item For $k\in\{1,2, \dots, D\}$ and $z^{L,k}\in\R_+,$ 
    limit buy orders arrive $k$ ticks below the best ask price with intensity $z^{L,k}\bI^+_t,$ and limit sell orders arrive $k$ ticks above the best bid price with intensity $z^{L,k}\bI^-_t.$ 
	\item If the spread is $j$ ticks, then for $k\in\{j,\dots,D\}$ and $z^{C,k}\in\R_+,$ cancellations of limit buy orders arrive $k$ ticks below the best ask price with intensity $z^{C,k}\bI^-_t v^{b,{k-j+1}}$, and cancellations of limit sell orders arrive
    $k$ ticks above the best bid price with intensity $z^{C,k}\bI^+_t v^{a,{k-j+1}}.$ 
\end{itemize}

The introduction of tactical and strategic traders increases the difficulty of the market-making problem. In markets containing only noise traders, order flow is largely symmetric and not directional. When tactical and strategic traders are present, the market exhibits directional pressure. This exposes the market maker to the risk of limit orders being filled immediately before an unfavorable price move, a phenomenon known as {\sl adverse selection}, which we study in more detail in Section~\ref{sec:markouts}.

\subsection{Order sizes}
\label{sec:order_size}
The order sizes of noise, tactical, and strategic traders follow a half-normal distribution. Let $Z$ be a random variable with standard normal distribution. For $k\in\{1,\dots, D\},$ market orders, limit orders, and cancellations of the noise traders arrive with sizes $1+\traderstd^{M}_\text{noise}|Z|, 1+\traderstd^{L,k}_\text{noise}|Z|,$ and $1+\traderstd^{C,k}_\text{noise}|Z|.$ Orders of the tactical traders arrive with sizes $1+\traderstd^{M}_\text{tactical}|Z|, 1+\traderstd^{L,k}_\text{tactical}|Z|$ and $1+\traderstd^{C,k}_\text{tactical}|Z|.$ Finally, orders of the strategic traders arrive with sizes $1+\traderstd^{M}_\text{strategic}|Z|, 1+\traderstd^{L,k}_\text{strategic}|Z|$ and $1+\traderstd^{C,k}_\text{strategic}|Z|.$ Order sizes are always rounded to the closest integer, and we draw a separate realization of $Z$ for each order arrival. All types of traders cancel orders with the highest queue positions first and cannot cancel more orders than are currently resting on that price level.

\section{Numerical experiments}
\label{sec:numerical_experiments}

We compare the logistic-normal (LN) actor-critic algorithm with three heuristic benchmark algorithms in three market environments consisting of the traders described in Section~\ref{sec:market_simulation_details}. In this section, we describe all aspects of the numerical experiments in detail. 

\subsection{Benchmark algorithms}
\label{sec:benchmark_algorithms}

We compare the performance of the LN algorithm against the following heuristic benchmarks, which place limit orders at times $t_n$ for $n=0,1,\dots,N-1.$
\begin{itemize}
    \item The TOP1 algorithm places $M/2$ lots on the best bid and the best ask at each time $t_n.$
    \item The TOP2 algorithm places $M/2$ lots one tick below the best bid price and one tick above the best ask price at each time $t_n.$
    \item The INV algorithm places orders at the best prices but skews the order sizes as a function of current inventory. Let $\inventory_n$ be the inventory at time $t_n$ and define
    \begin{equation}
        \bar{\inventory}_n =
        \max \left(\min \left(\frac{\alpha \inventory_n}{M}, 1\right),-1\right),
        \label{eq:inventory_skew}
    \end{equation}
    for a parameter $\alpha>0,$ which controls the aggressiveness of the algorithm's response to inventory deviations. Then, the algorithm posts $\frac{M}{2}(1-\bar{\inventory}_n)$ lots at the best bid price and $\frac{M}{2}(1+\bar{\inventory}_n)$ lots at the best ask price at each time $t_n.$ We round volumes to integers with the method described in Section~\ref{sec:action_space}.
\end{itemize}
To mirror the terminal reward in \eqref{eq:terminal_reward}, the benchmark algorithms send a market order at terminal time $t_N$ that reduces their absolute inventory to at most $\ceil{\nu M},$ generating the cash flow $\text{MO}_\nu(s_N)$. Furthermore, the algorithms only cancel resting orders when necessary to place new ones, in order to retain queue positions.

\subsection{Simulation setup}

We test all algorithms in three different market environments: one with only noise traders, one with noise and tactical traders, and one with noise, tactical, and strategic traders. The algorithms make decisions at intervals of $\Delta t = 30\,\mathrm{s}$ until the terminal time $T = 600\,\mathrm{s},$ for a total of $N = 20$ decision steps. At terminal time $t_N=T,$ the algorithms must liquidate their entire inventory by sending a market order, which corresponds to $\nu=0$ in \eqref{eq:terminal_reward}. In Appendix~\ref{sec:effect_of_terminal_inventory_constraint}, we also test the performance of the algorithm when the terminal inventory constraint is relaxed to $\nu=0.5.$ 

\begin{figure}[htbp]
    \centering
    \includegraphics[width=0.75\linewidth]{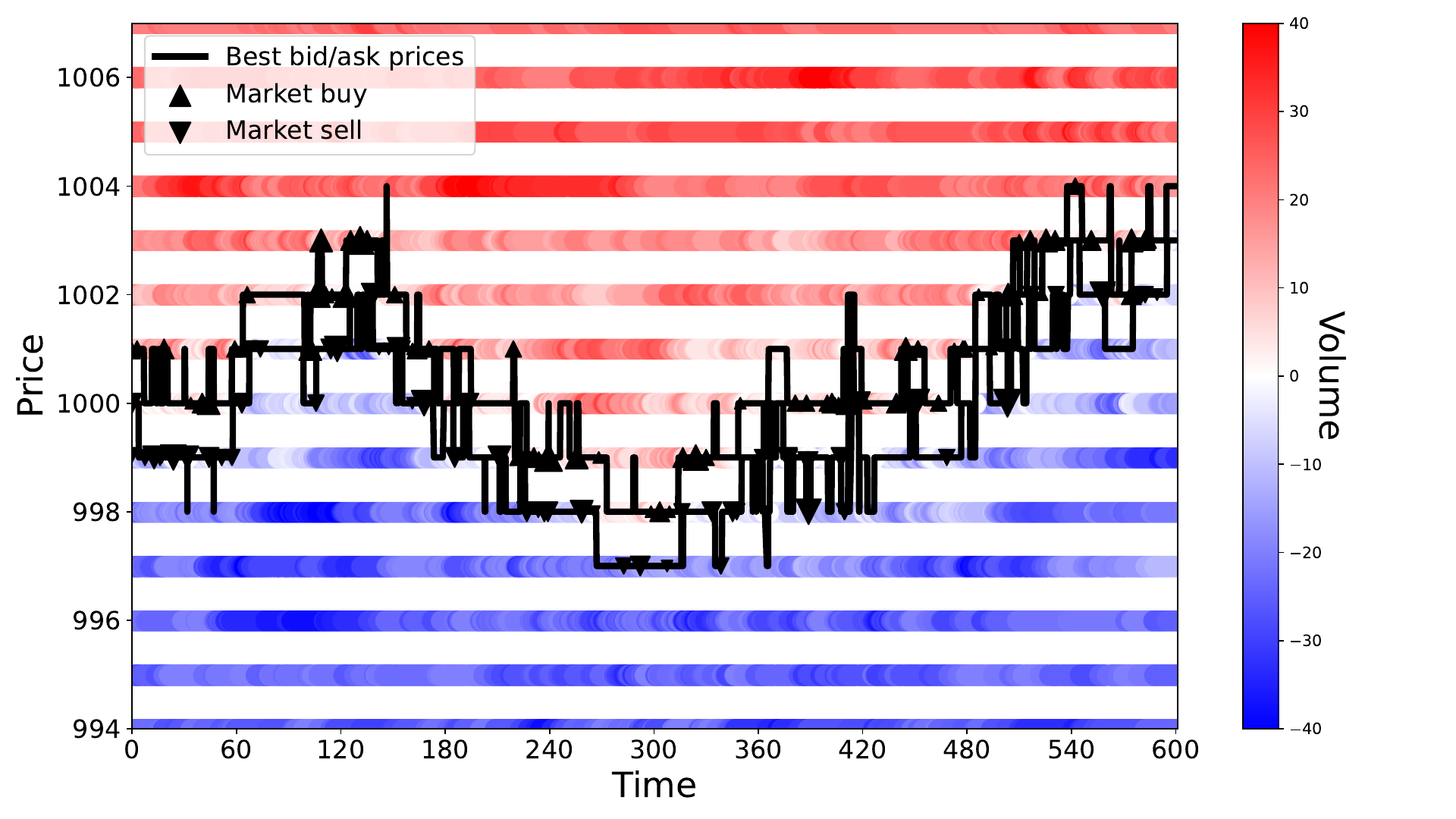}
    \caption{Evolution of the order book over time. Red colors indicate limit sell orders, while blue colors indicate limit buy orders. The black lines represent the best bid and ask prices. The triangles indicate market buy and sell orders. Larger sizes of the triangles correspond to larger order sizes. Deeper color tones correspond to larger volumes.}
    \label{fig:heat_map}
\end{figure}

To illustrate the market simulation, Figure~\ref{fig:heat_map} shows the evolution of the order book from time $t=0\,\mathrm{s}$ until time $t=600\,\mathrm{s}$ for the market consisting solely of noise traders. 
All simulation parameters and implementation details are provided in Appendix~\ref{sec:simulation_parameters}.

\subsection{Markouts}
\label{sec:markouts}
Before comparing the performance of the algorithms, we first analyze the market environments in more detail. The tactical traders only react to the instantaneous volume imbalance $I_t$. Because of the order-book shape shown in Figure~\ref{fig:shape}, the imbalance tends to mean-revert after a price level is removed. Therefore, the tactical traders trade in short directional bursts. In contrast, because the strategic traders observe a smoothed imbalance signal, their directional trades are longer-lasting, leading to sustained price moves. In a market populated by both tactical and strategic traders, market making becomes more challenging. Passive limit orders may be executed just before a price move triggered by tactical or strategic traders, exposing the market maker to adverse selection.

To evaluate the degree of adverse selection across environments, we compute the markouts of the limit orders for the TOP1 and TOP2 algorithms when they place $M=2$ or $M=20$ lots in the book. A markout of a limit order fill at time $t$ at price $p^f_t$ is defined by 
\begin{equation}
\text{markout}_t =
\begin{cases}
    p^f_t - p_{t+\Delta t} \quad \text{if limit sell}, \\ 
    p_{t+\Delta t} - p^f_t \quad \text{if limit buy}, 
\end{cases}
\label{eq:markout}
\end{equation}
where $\Delta t>0$ is some time interval and $p_t$ is the mid-price. Markouts measure the extent of adverse selection for limit order fills. A limit sell order is adversely selected if the mid-price rises after the order is filled, so that a better selling price could have been achieved later. A limit buy order is adversely selected if the mid-price moves lower after the order is filled, so that a lower buying price could have been achieved later.

\begin{figure}[htbp]
	\centering
	\begin{subfigure}[t]{0.33\textwidth}
    \includegraphics[width=\textwidth]{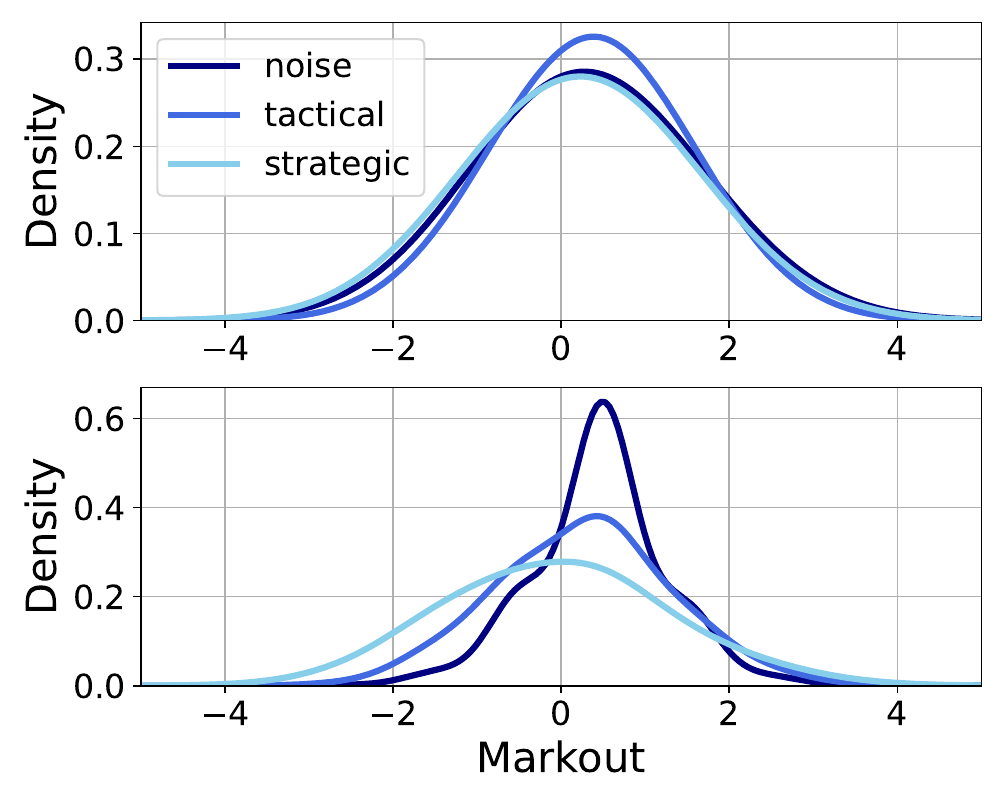}
    \caption{Markouts for TOP1.}
	\end{subfigure}
	\hspace{1em}
    \begin{subfigure}[t]{0.33\textwidth}
    \includegraphics[width=\textwidth]{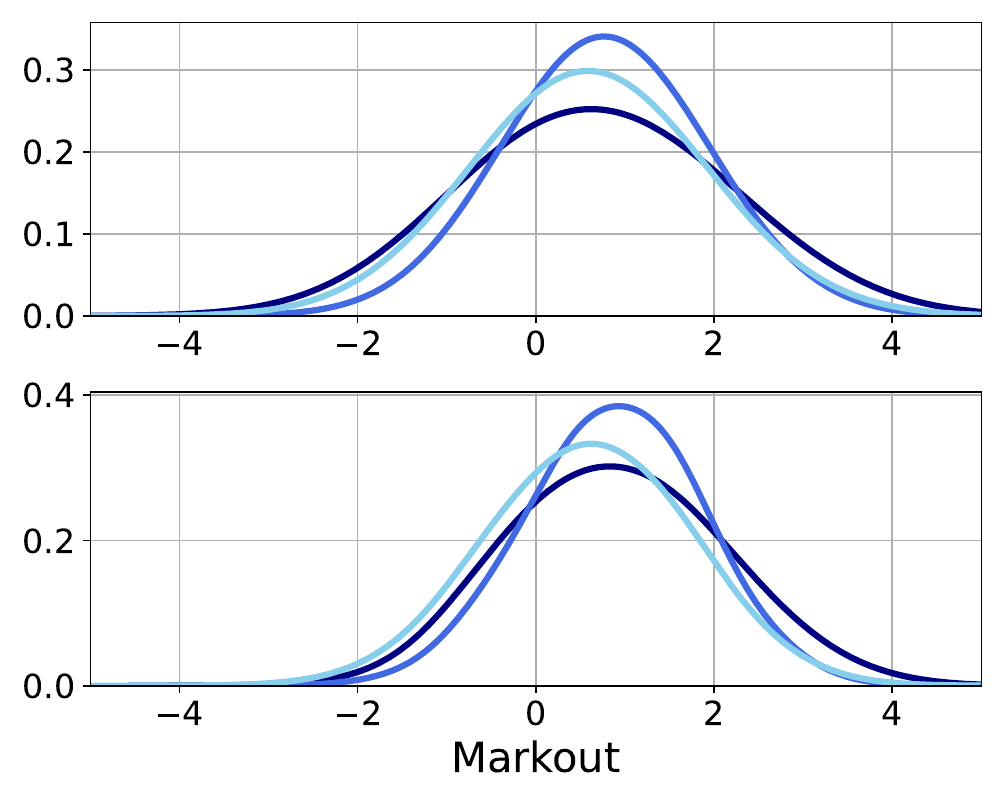}
    \caption{Markouts for TOP2.}
	\end{subfigure} 
	\caption{The upper row corresponds to markouts of the algorithms posting $M=2$ lots. The lower row corresponds to markouts of the algorithms posting $M=20$ lots. Color shades correspond to market environments. Dark-blue corresponds to the noise market, mid-blue corresponds to the market consisting of noise and tactical traders, and light-blue corresponds to the market consisting of noise, tactical, and strategic traders.}
    \label{fig:markouts}
\end{figure}

Here, we choose $\Delta t = 30\,\mathrm{s}$ in line with the decision frequency of the market-making algorithms. In our simulations, the spread is usually one tick. Therefore, expected markouts of limit order fills at the best prices are around 0.5 ticks in an environment where prices are stable with few adverse selection effects. Table~\ref{table:markouts} and Figure~\ref{fig:markouts} summarize expected markouts and histograms of markouts for the TOP1 and TOP2 algorithms, placing $M=2$ or $M=20$ lots. The markout statistics are based on 10,000 limit order fills per algorithm, number of lots, and market environment. We deliberately restrict this analysis to the benchmark algorithms, whose placement is fixed and identical across the three environments, which isolates the effect of the environment on adverse selection. The LN algorithm adapts its placement to each environment, as we show in Section~\ref{sec:action_and_volume_analysis}, so that its markouts would confound the properties of the market with the response of the policy.

We observe that markouts for the TOP2 algorithm are better than for the TOP1 algorithm. This is clear from the definition \eqref{eq:markout}, as the distance from the current mid-price $p_t$ at fill time is higher than for the TOP1 algorithm. However, limit fills deeper in the book are also less likely. So, a higher markout per limit order does not necessarily mean the algorithm will be more profitable, since those fills are less frequent. We observe that the markouts for both algorithms deteriorate once strategic traders are introduced, meaning that the orders are adversely selected more often. It is not always the case that placing more lots leads to better markouts. For the TOP1 algorithm, markouts of the algorithm placing 20 lots are better than when it is placing two lots in the market with noise traders. But in all other markets, the markouts when it is placing 20 lots are worse because placing more lots also means that more orders are affected by adverse price moves. The same effect does not hold for the TOP2 algorithm, as placing at the second-best prices shields the limit order fills against adverse price moves. 

\begin{table}[htbp]
\begin{center}
\begin{scriptsize}
\begin{sc}
\begin{tabular}{lccccc}
\toprule
 & Lots & $\E[\text{MO-TOP1}]$ & $\sigma[\text{MO-TOP1}]$ & $\E[\text{MO-TOP2}]$ & $\sigma[\text{MO-TOP2}]$ \\
\midrule
noise & 2 & 0.33 & 1.17 & 0.68 & 1.21 \\
 & 20 & 0.45 & 0.75 & 0.90 & 1.07 \\
noise \& tactical & 2 & 0.36 & 1.03 & 0.78 & 0.92 \\
 & 20 & 0.26 & 0.99 & 0.88 & 0.86 \\
noise \& tactical & 2 & 0.23 & 1.18 & 0.60 & 1.05 \\
\& strategic & 20 & $-0.06$ & 1.27 & 0.59 & 1.00 \\
\bottomrule
\end{tabular}

\end{sc}
\end{scriptsize}
\end{center}
\caption{Expected value and standard deviation of the markouts defined in \eqref{eq:markout} for the TOP1 and TOP2 algorithms' limit order fills, and for $M=2$ and $M=20$ lots.
}
\label{table:markouts}
\end{table}

\subsection{Evaluation and reinforcement learning setup}
\label{sec:evaluation_and_rl_setup}
\begin{figure}[htbp]
	\centering
	\begin{subfigure}[t]{0.3\textwidth}
		\includegraphics[width=\textwidth]{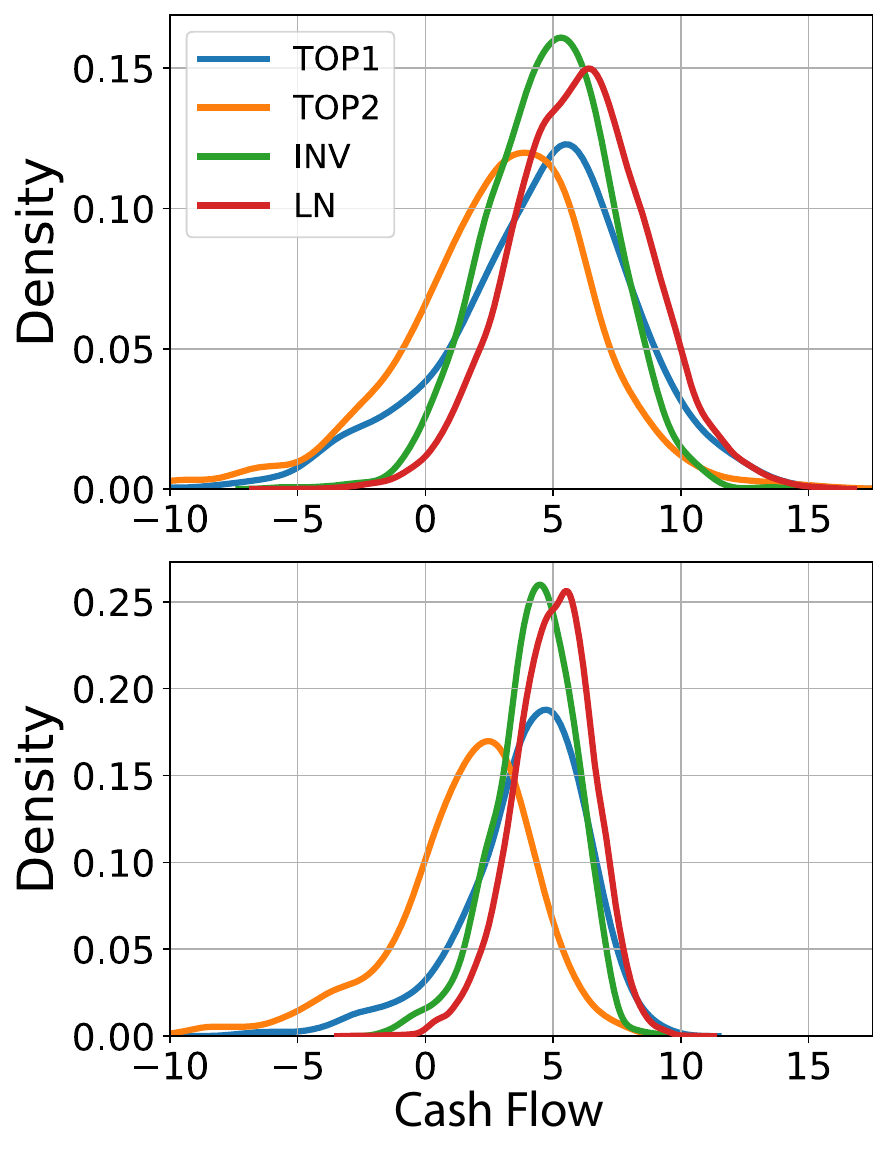}
		\caption{Noise}
		\label{fig:noise_results}
	\end{subfigure}
	\hfill
	\begin{subfigure}[t]{0.3\textwidth}
		\includegraphics[width=\textwidth]{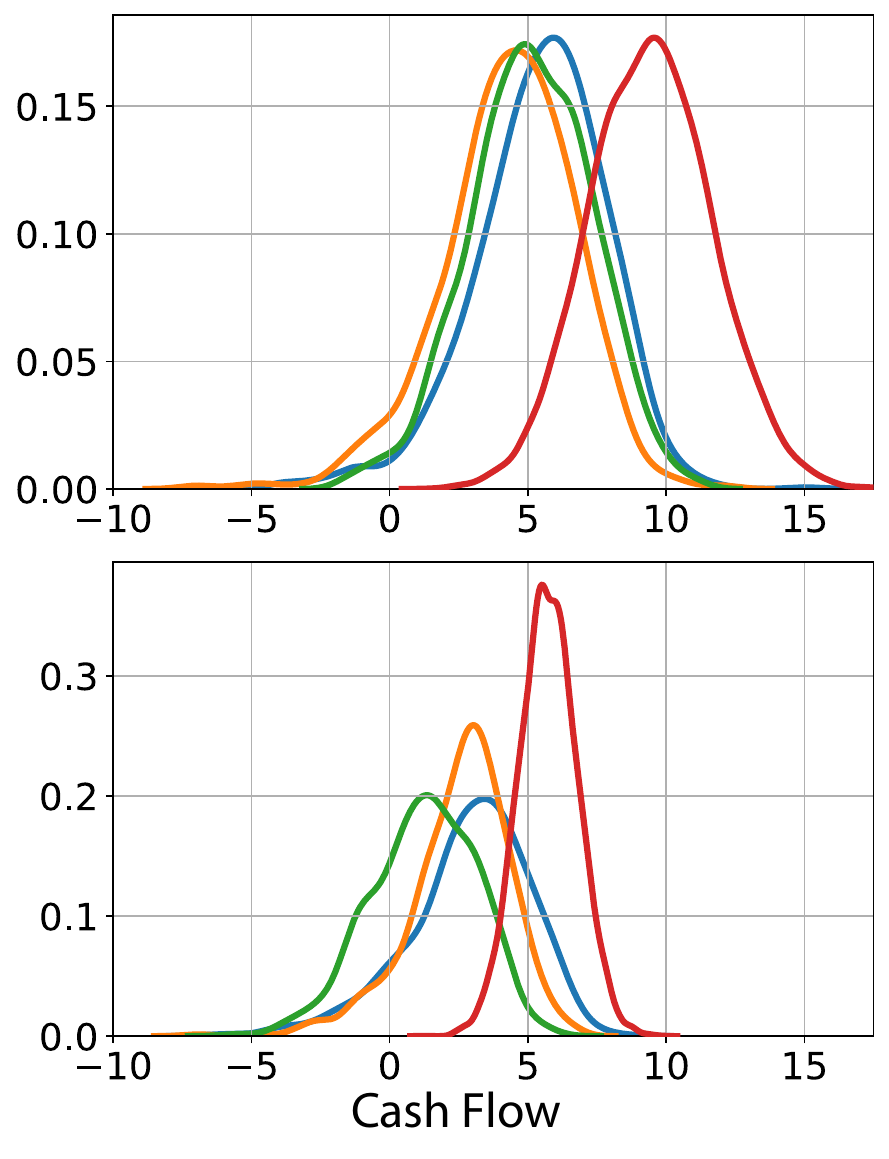}
		\caption{Noise \& Tactical}
		\label{fig:tactical_results}
	\end{subfigure}
	\hfill 
	\begin{subfigure}[t]{0.3\textwidth}
	\includegraphics[width=\textwidth]{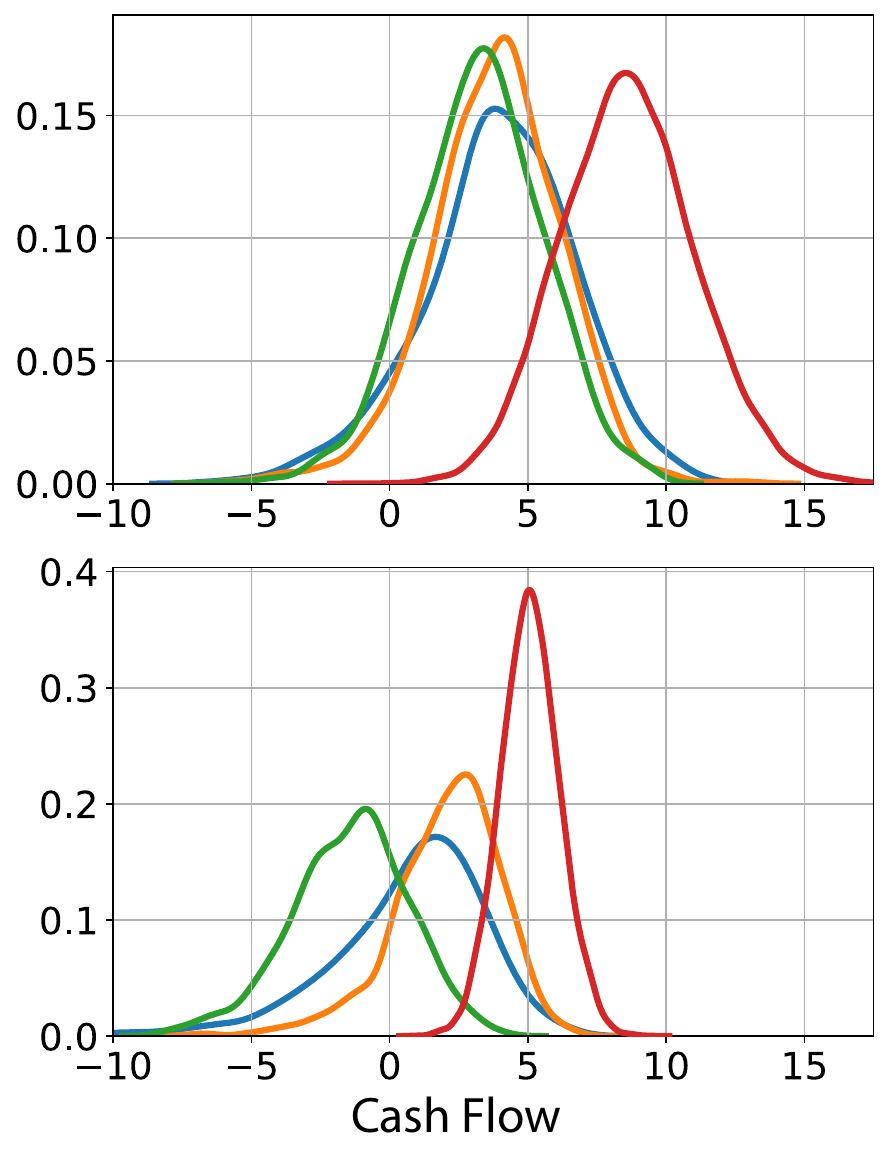}
	\caption{Noise, Tactical \& Strategic}
	\label{fig:strategic_results}			
	\end{subfigure}	
	\caption{Histograms of the normalized cash flow \eqref{eq:sum_of_rewards_normalized} for the TOP1, TOP2, INV, and LN algorithms, for all markets, and for $M=2$ and $M=20$ lots. }
	\label{fig:histogram}
\end{figure}

We compare the LN algorithm with the three heuristic benchmark algorithms TOP1, TOP2, and INV, where we set the inventory sensitivity parameter of the INV algorithm in \eqref{eq:inventory_skew} to $\alpha=1.0.$ We assess each algorithm's performance by computing the following normalized cash flow
\begin{equation} 
\frac{1}{M}\left( \sum_{n=0}^{N-1} \bar{r}(s_n, \action_n) + \inventory_{N+}p_N + \text{MO}_\nu(s_N) \right)
,
\label{eq:sum_of_rewards_normalized}
\end{equation}
where $\bar{r}(s_n, \action_n)$ are the cash flows from market and limit order fills and $\text{MO}_\nu(s_N)$ is the cash flow from the terminal market order, as defined in Section~\ref{sec:rewards}. In contrast to the training objective \eqref{eq:telescoping}, the metric \eqref{eq:sum_of_rewards_normalized} excludes the running inventory penalty $-\gamma\lvert \inventory_{n+1}\rvert,$ so that the reported cash flows remain comparable across different values of $\gamma.$ Table~\ref{table:pnl} shows the empirical expected values and standard deviations of \eqref{eq:sum_of_rewards_normalized}, and Figure~\ref{fig:histogram} shows the histograms of \eqref{eq:sum_of_rewards_normalized}. The values are based on 10,000 test samples. We use separate random seeds for training and testing, to evaluate on out-of-sample episodes. In Appendix~\ref{sec:dirichlet_vs_logistic_normal}, we compare the performance of the logistic-normal distribution with a Dirichlet distribution, which also has support on the simplex, and find that the logistic-normal distribution performs better than the Dirichlet distribution.

For the LN algorithm, we choose an inventory risk parameter $\gamma=0.01$ in \eqref{eq:rewards} and $K=3$ for the dimension of the simplex $\SI^{2(K+1)}$. In Appendix~\ref{sec:effect_of_inventory_penalty}, we also conduct an experiment with $\gamma=0,$ which corresponds to no inventory penalty. The encoder network $f^o_\phi,$ mean network $f^m_{\tm},$ and the value network $f^V_{\vartheta}$ are implemented as standard feed-forward neural networks. We initialize the bias of the output layer of the network $f^m_{\tm}$ to $b=(1,1,\dots,1)\in\R^{2(K+1)}$ and set the remaining weights of this layer close to zero. Then, we obtain from \eqref{eq:mean} that
\begin{equation}
	\label{eq:bias}
	\E \left[ \log\left( \frac{\action^k}{\action^0} \right) \right] \approx 1, 
	\quad 
    k=1,2,\dots,2(K+1).
\end{equation}
Thus, action $\action^0$ is initially less likely than actions $\action^1,\action^2, \dots, \action^{2(K+1)},$ which reduces the risk of early inactivity and of converging to a degenerate local optimum in which the algorithm never trades. The log-variance parameters are initialized as $(\theta^{v,1}, \dots, \theta^{v,2(K+1)})=(0,0,\dots,0)\in\R^{2(K+1)}.$ After training, we evaluate the algorithm by sampling actions $\action\sim\pi_{\phi, \theta}(\cdot\mid s),$ where $\phi$ and $\theta$ are the trained weights. Further implementation details are provided in Appendix~\ref{sec:rl_parameters}.

\subsection{Market with noise traders}

The results for the market with noise traders are displayed in the first two rows of Table~\ref{table:pnl}. In this environment, market and limit order arrivals do not depend on the state of the book; only the cancellation intensities scale linearly with the resting volume. We observe that the TOP1 algorithm achieves higher expected cash flows than TOP2, with both algorithms showing similar standard deviations, for both $M=2$ and $M=20.$ The INV algorithm performs better than the TOP1 algorithm, achieving similar expected cash flows but with lower standard deviations, indicating a positive effect of the inventory skewing. The LN algorithm outperforms the INV algorithm, with higher expected values and comparable standard deviations. 

\begin{table}[htbp]
\begin{center}
    \begin{scriptsize}
        \begin{sc}
\begin{tabular}{lccccccccc}
\toprule
 & Lots & $\E[\text{TOP1}]$ & $\sigma[\text{TOP1}]$ & $\E[\text{TOP2}]$ & $\sigma[\text{TOP2}]$ & $\E[\text{INV}]$ & $\sigma[\text{INV}]$ & $\E[\text{LN}]$ & $\sigma[\text{LN}]$ \\
\midrule
noise & 2 & 4.40 & 3.87 & 2.91 & 3.69 & 4.83 & 2.40 & \textbf{6.03} & 2.81 \\
 & 20 & 3.84 & 2.62 & 1.48 & 2.82 & 4.26 & 1.59 & \textbf{4.66} & 1.41 \\
noise \& tactical & 2 & 5.47 & 2.38 & 4.31 & 2.42 & 5.11 & 2.20 & \textbf{9.11} & 2.20 \\
 & 20 & 2.88 & 2.21 & 2.55 & 1.82 & 1.22 & 1.92 & \textbf{5.68} & 1.03 \\
noise \& tactical & 2 & 3.95 & 2.76 & 3.83 & 2.35 & 3.24 & 2.32 & \textbf{8.55} & 2.40 \\
\& strategic & 20 & 0.73 & 2.73 & 2.05 & 1.95 & $-1.44$ & 2.12 & \textbf{4.99} & 1.07 \\
\bottomrule
\end{tabular}
        \end{sc}
    \end{scriptsize}
\end{center}
\caption{Expected value and standard deviation of the normalized cash flow \eqref{eq:sum_of_rewards_normalized} for the TOP1, TOP2, INV, and LN algorithms, for all markets, and for $M=2$ and $M=20$ lots.
}
\label{table:pnl}
\end{table}

\subsection{Market with noise \& tactical traders}

The results are summarized in the third and fourth rows of Table~\ref{table:pnl}. The performance of the TOP2 algorithm is now closer to the performance of the TOP1 algorithm, because tactical traders induce price jumps. In this environment, the TOP1 algorithm has the best expected cash flows among the benchmark algorithms, although with higher standard deviations. For $M=2$ lots, the INV algorithm has the second-best performance, with lower standard deviations than TOP1. For $M=20$ lots, it has the worst performance. This indicates that simple inventory skewing is no longer effective in this market, likely because inventory rebalancing occurs more frequently during adverse price moves. This suggests that inventory control must be complemented with predictive signals derived from the limit order book. Consistent with this interpretation, the performance of the LN algorithm is much stronger than that of the benchmark algorithms, with higher expected cash flows and lower standard deviations, as it integrates inventory risk management with predictive order-book signals.

\subsection{Market with noise \& tactical \& strategic traders}

When the strategic traders are present in the market, order flow becomes both state-dependent and persistent, leading to pronounced price trends that drive adverse selection and increase inventory risk for liquidity providers. This is also visible in the lower markouts in the fifth and sixth rows of Table~\ref{table:markouts}. For $M=2$ lots, the performance of the TOP2 algorithm is close to the performance of the TOP1 algorithm, measured in expected value and standard deviation. The expected cash flows of the INV algorithm are the worst, although with the lowest standard deviation. For $M=20$ lots, the TOP2 algorithm performs best because it is less likely to be picked off by adverse fills on the first level, to which the TOP1 algorithm is more vulnerable. The expected cash flows of the INV algorithm are the worst among the three benchmarks, with lower standard deviations than TOP1 and higher standard deviations than TOP2. The poor performance is likely driven by adverse executions when skewing inventory.
The LN algorithm performs best with high expected cash flows and low standard deviations, particularly for $M=20$. The performance gap between LN and the benchmarks is largest in this environment, demonstrating the benefit of adaptive, state-dependent policies that integrate inventory control with predictive order-book signals when adverse selection effects are most severe.

\section{Analyzing trading behavior}
\label{sec:action_and_volume_analysis}

In this section, we analyze the algorithm's trading behavior in more detail, by studying its actions, order placement, and inventory management. As in Section~\ref{sec:numerical_experiments},
the results are based on the same 10,000 test episodes. Furthermore, the simulation and the LN algorithm are configured with the same hyperparameters as in Section~\ref{sec:numerical_experiments}.

\subsection{Actions}

\begin{figure}[htbp]
	\centering
	\begin{subfigure}[t]{0.3\textwidth}
		\includegraphics[width=\textwidth]{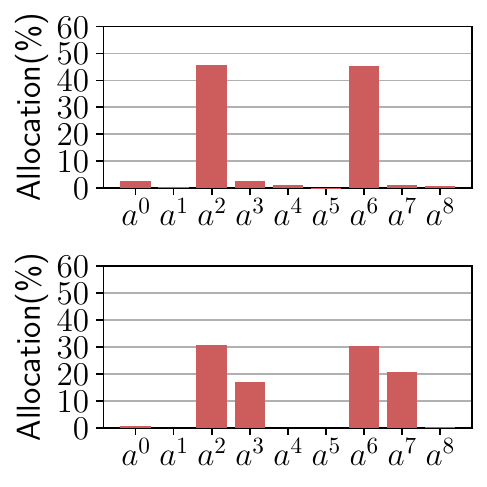}
		\caption{Noise}
		\label{fig:actions_noise}
	\end{subfigure}
	\hfill
	\begin{subfigure}[t]{0.3\textwidth}
		\includegraphics[width=\textwidth]{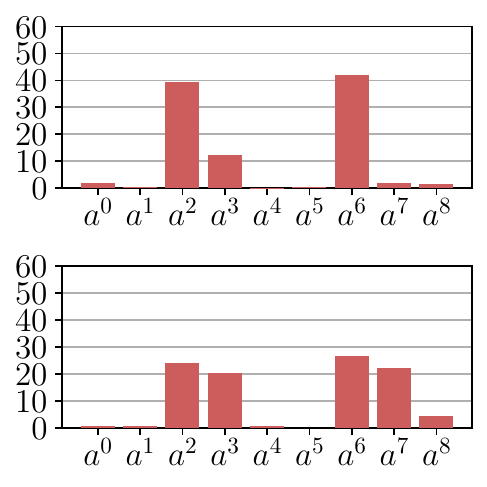}
		\caption{Noise \& Tactical}
		\label{fig:actions_tactical}
	\end{subfigure}
	\hfill 
	\begin{subfigure}[t]{0.3\textwidth}
	\includegraphics[width=\textwidth]{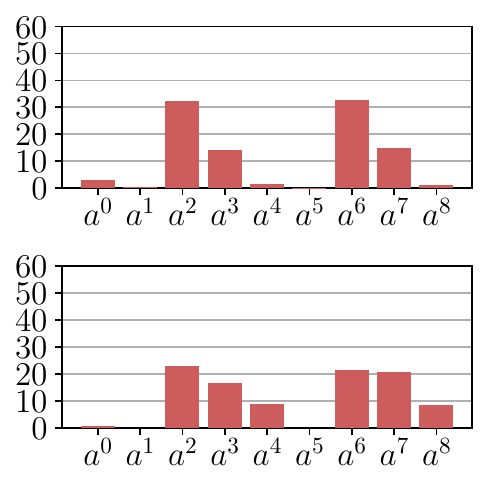}
	\caption{Noise, Tactical \& Strategic}
	\label{fig:actions_strategic}			
	\end{subfigure}	
	\caption{Average action allocations of the LN algorithm, averaged over all time steps and episodes. Actions for $M=2$ lots correspond to the upper panel, while actions for $M=20$ lots correspond to the lower panel. The columns represent different market environments.}
	\label{fig:actions}
\end{figure}

Figure~\ref{fig:actions} shows a bar plot of the average components of the actions. The bars correspond to actions averaged over all time steps and episodes. The columns correspond to different market environments. The upper panel shows $M=2$ lots, while the lower panel shows $M=20$ lots. We recall that the action $\action^0$ denotes the fraction of the total number of lots $M$ that is not placed in the book, while the actions $\action^1, \action^2, \action^3, \action^4$ correspond to the fractions of $M$ allocated to market buy and limit buy orders, and the actions $\action^5, \action^6, \action^7, \action^8$ correspond to the fractions of $M$ allocated to market sell and limit sell orders.

Across all markets and both choices of $M$, the allocations to inactivity $\left(\action^0\right)$, market buy $\left(\action^1\right)$, and market sell $\left(\action^5\right)$ are negligible, and the algorithm's actions mostly correspond to limit order placements. Furthermore, we notice that the action distributions are not completely symmetric across all markets and for $M=2$ and $M=20.$ The buy and the sell side receive nearly the same total allocation, but the split across price levels differs slightly, as for instance in the market with noise and tactical traders and $M=2$ lots, where the second-best price is used on the bid side but hardly on the ask side. These residual asymmetries do not lead to one-sided trading: as shown in Section~\ref{sec:inventory_evolution}, the algorithm maintains a balanced inventory over time.

For the market consisting of noise traders and for $M=2$ lots, we observe that the volume is placed almost exclusively at the best bid and ask prices, using actions $\action^2$ and $\action^6,$ while for $M=20$ lots the algorithm splits its volume between the best and the second-best prices, using in addition the actions $\action^3$ and $\action^7.$ 

In the market consisting of noise and tactical traders, the algorithm starts to use the second-best bid price already for $M=2$ lots, and for $M=20$ lots the allocations to the best and second-best prices are of comparable size, with a small allocation to the third price level, mainly through the action $\action^8$.

Once the strategic traders are added, the shift away from the best prices is most pronounced. For $M=2$ lots, a visible share of the volume is placed at the second-best prices, and for $M=20$ lots the volume is distributed over all three price levels, with a clearly larger allocation to the third-best levels than in the previous market. This progression is consistent with the increasing degree of adverse selection. In the market with noise, tactical, and strategic traders, adverse price moves become more persistent, making orders at the top of the book more exposed to being picked off right before a price move. 

\begin{figure}[htbp]
	\centering
	\begin{subfigure}[t]{0.3\textwidth}
		\includegraphics[width=\textwidth]{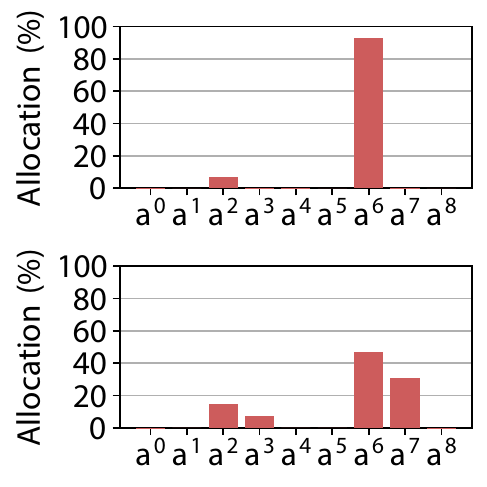}
		\caption{Noise}
		\label{fig:actions_imb_noise}
	\end{subfigure}
	\hfill
	\begin{subfigure}[t]{0.3\textwidth}
		\includegraphics[width=\textwidth]{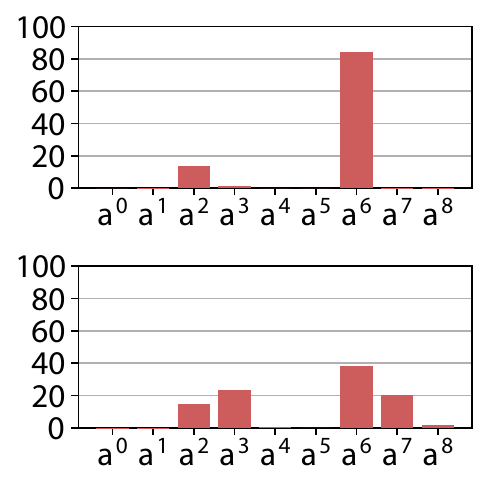}
		\caption{Noise \& Tactical}
		\label{fig:actions_imb_tactical}
	\end{subfigure}
	\hfill 
	\begin{subfigure}[t]{0.3\textwidth}
	\includegraphics[width=\textwidth]{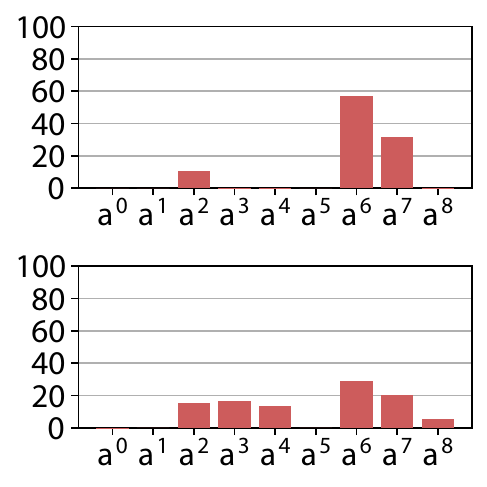}
	\caption{Noise, Tactical \& Strategic}
	\label{fig:actions_imb_strategic}			
	\end{subfigure}	
	\caption{Average action allocations of the LN algorithm, conditional on $\inventory_n\geq M/2,$ averaged over all time steps and episodes satisfying this condition. Actions for $M=2$ lots correspond to the upper panel, while actions for $M=20$ lots correspond to the lower panel. The columns represent different market environments.}
	\label{fig:actions_imb}
\end{figure}

In Figure~\ref{fig:actions_imb}, we evaluate the algorithm's average actions at all time steps $t_n$ conditional on $\inventory_n \geq M/2.$ In these states, the algorithm carries a substantial one-sided inventory, and we expect it to place more sell orders, as a large inventory contributes negatively to the reward function in \eqref{eq:rewards}. Indeed, for $M=2$ lots, the placement behavior is strongly asymmetric across all markets, with the large majority of the volume allocated to the sell side through the actions $\action^6$ and $\action^7,$ and only a small residual allocation to the best bid. For $M=20$ lots, the behavior is more balanced, but the ask side of the book still receives the larger share of the volume in all three markets. The comparison across markets also shows that the limit order skewing is achieved in different ways. In the markets with noise and with noise and tactical traders, the sell volume is concentrated at the best ask, whereas in the market with strategic traders it is spread more evenly across order-book levels.

\subsection{Limit orders, cancellations, and market orders}

In Table~\ref{table:fills_by_level}, we analyze the limit orders, cancellations, and market orders of the LN algorithm in more detail. The table consists of three blocks, reporting the limit order fills, the limit order cancellations, and the market orders. The column \textsc{Rate} of the fill block reports the share of the limit order volume placed in the book that is eventually filled, the column \textsc{Rate} of the cancellation block reports the share of limit order volume that is cancelled, and the column \textsc{Rate} of the market order block reports the share of the algorithm's total placed volume that is executed with market orders. This share includes the market order that liquidates the remaining inventory at the terminal time, which is imposed by the terminal inventory constraint in \eqref{eq:terminal_reward} with $\nu=0.$ Every placed order is eventually either filled or cancelled, so that the limit fill and cancellation rates add up to $100\%.$ Within each block, the columns $L^1$, $L^2$, and $L^3$ decompose the corresponding volume by the price level at which the order was inserted: $L^1$ corresponds to the best price, $L^2$ to the second-best price, and $L^3$ to the third-best price. We report the insertion level rather than the level at the time of the fill, because limit orders are filled almost exclusively at the best price, so that the level at fill time would carry little information. For the cancellations, the two conventions do not differ, since the algorithm cancels its resting orders immediately after making a trading decision.

The table refines the picture obtained from the action allocations in Figure~\ref{fig:actions}, which show how the algorithm distributes its volume across the price levels. The level columns of the fill block instead show how the volume that is eventually filled was distributed at insertion. This distribution is tilted towards the best prices, because orders inserted there are filled with a higher probability.

In the market with noise traders and $M=2$ lots, essentially the entire filled volume was inserted at the best prices, whereas for $M=20$ lots a quarter of it was inserted at the second-best prices. Adding the tactical traders moves the insertion level of the filled volume further away from the top of the book, and the shift is most pronounced when the strategic traders are present as well. In that market, roughly a quarter of the filled volume was inserted at the second-best prices for $M=2$ lots, and $8.30\%$ was inserted at the third-best prices for $M=20$ lots, compared to $2.10\%$ in the market with noise and tactical traders. The cancelled volume shows the same shift towards deeper insertion levels across the three markets. In addition, for $M=20$ lots it sits at deeper levels than the filled volume, which is expected, since orders inserted at the best prices are more likely to be filled before they are cancelled.

The fill rate decreases as tactical and strategic traders enter the market, from $74.65\%$ to $59.16\%$ for $M=2$ lots and from $59.65\%$ to $43.99\%$ for $M=20$ lots, and it decreases likewise when the algorithm quotes more lots. Because fill and cancellation rates add up to $100\%,$ the cancellation rate increases when tactical and strategic traders are present, and reaches $56.01\%$ in the market with strategic traders for $M=20$ lots. Cancellations are necessary to move resting volume to a different price level, so this reflects a more active management of resting orders in the markets where the order flow is more informed.

The algorithm barely uses market orders by its own choice, in line with the negligible allocations to the actions $\action^1$ and $\action^5$ in Figure~\ref{fig:actions}. This is because market orders require crossing the spread, whereas limit orders earn it. The share reported in the market order block lies between $1.90\%$ and $2.74\%$ across all markets and both lot sizes, and it is driven almost entirely by the terminal market order, which is imposed by the terminal inventory constraint.

\begin{table}[htbp]
\begin{center}
    \begin{scriptsize}
        \begin{sc}
\begin{tabular}{lcccccccccc}
\toprule
 & & \multicolumn{4}{c}{Limit order fills (\%)} & \multicolumn{4}{c}{Limit order cancellations (\%)} & \multicolumn{1}{c}{Market orders (\%)} \\
\cmidrule(lr){3-6} \cmidrule(lr){7-10} \cmidrule(lr){11-11}
 & Lots & Rate & $L^1$ & $L^2$ & $L^3$ & Rate & $L^1$ & $L^2$ & $L^3$ & Rate \\
\midrule
noise & 2 & 74.65 & 99.71 & 0.29 & 0.01 & 25.35 & 97.96 & 1.80 & 0.24 & 2.67 \\
 & 20 & 59.65 & 74.64 & 25.36 & 0.00 & 40.35 & 49.29 & 50.68 & 0.04 & 2.10 \\
noise \& tactical & 2 & 63.93 & 91.76 & 8.24 & 0.00 & 36.07 & 88.76 & 10.76 & 0.48 & 2.73 \\
 & 20 & 52.88 & 65.35 & 32.55 & 2.10 & 47.12 & 49.15 & 45.20 & 5.65 & 2.60 \\
noise \& tactical & 2 & 59.16 & 73.39 & 26.59 & 0.02 & 40.84 & 75.10 & 24.03 & 0.87 & 2.74 \\
\& strategic & 20 & 43.99 & 61.80 & 29.89 & 8.30 & 56.01 & 42.15 & 35.33 & 22.52 & 1.90 \\
\bottomrule
\end{tabular}
        \end{sc}
    \end{scriptsize}
\end{center}
\caption{Limit order fills, cancellations, and market order executions of the LN algorithm, for all markets, and for $M=2$ and $M=20$ lots. In each of the two limit order blocks, the column \textsc{Rate} reports the share of the placed limit order volume that is filled, respectively cancelled, and the columns $L^1$, $L^2$, and $L^3$ decompose that volume into the shares that were inserted at the best, second-best, and third-best prices. The column \textsc{Rate} of the market order block reports the share of the total placed volume that is executed with market orders.}
\label{table:fills_by_level}
\end{table}

\subsection{Inventory evolution}
\label{sec:inventory_evolution}

\begin{figure}[htbp]
	\centering
	\begin{subfigure}[t]{0.3\textwidth}
		\includegraphics[width=\textwidth]{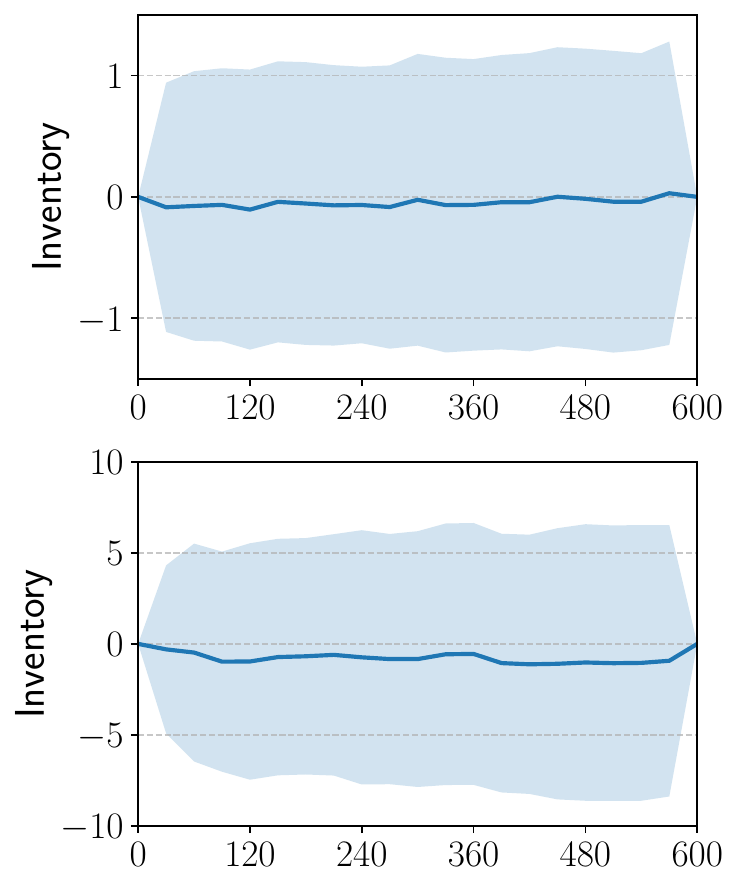} 
		\caption{Noise}
		\label{fig:inventory_noise}
	\end{subfigure}
	\hfill
	\begin{subfigure}[t]{0.3\textwidth}
		\includegraphics[width=\textwidth]{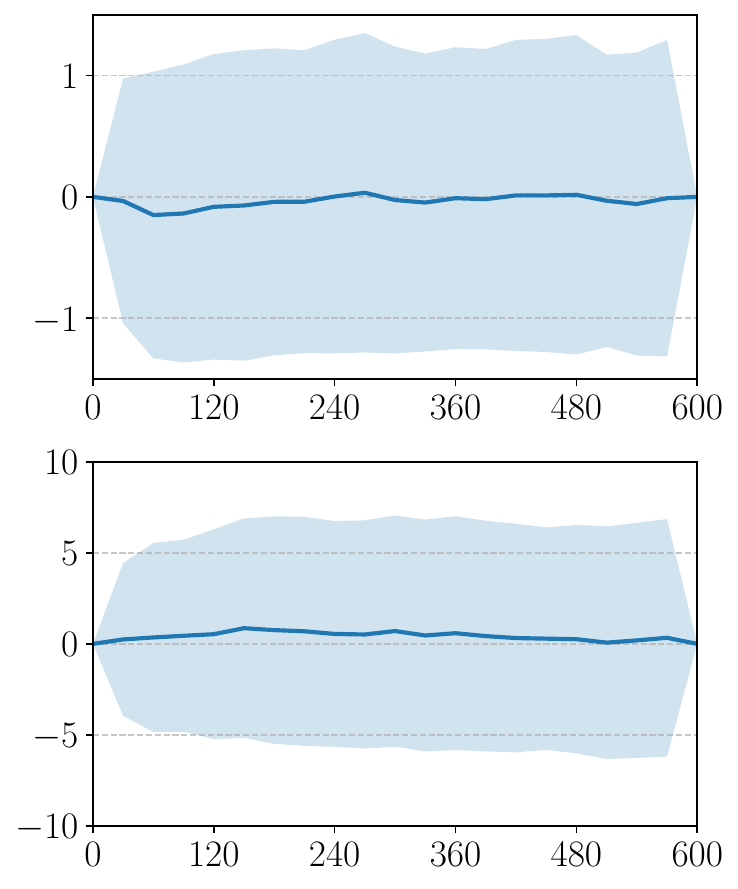}
		\caption{Noise \& Tactical}
		\label{fig:inventory_tactical}
	\end{subfigure}
	\hfill 
	\begin{subfigure}[t]{0.3\textwidth}
	\includegraphics[width=\textwidth]{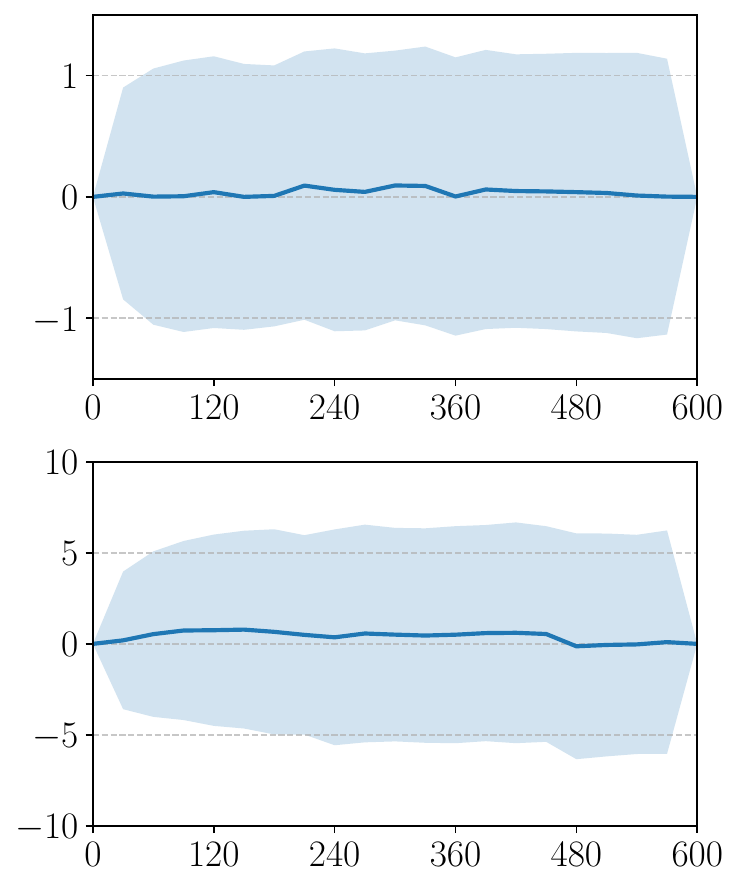}
	\caption{Noise, Tactical \& Strategic}
	\label{fig:inventory_strategic}			
	\end{subfigure}	
	\caption{Average inventory paths from time $t=0\,\mathrm{s}$ to time $t=600\,\mathrm{s}$ of the LN algorithm for three market environments. The upper panels correspond to $M=2$ lots, while the lower panels correspond to $M=20$ lots. The shaded regions indicate standard deviations. }
	\label{fig:inventory}
\end{figure}

Figure~\ref{fig:inventory} presents the average inventories across all episodes and market environments, with shaded regions indicating the corresponding standard deviations. In all three markets, and for both $M=2$ and $M=20$ lots, the average inventory stays essentially flat at zero over the entire episode, and the standard-deviation bands are almost symmetric around it. The remaining deviations of the mean from zero are an order of magnitude smaller than the width of the bands, so that we do not observe a systematic tilt towards long or short positions in any of the markets. The bands widen over the first few decision times, stay roughly constant during the episode, and contract at the terminal time, where the algorithm is forced to flatten its position, since we set $\nu=0$ in \eqref{eq:terminal_reward}. The width of the bands depends on the number of lots. For $M=2$ lots, the inventory stays within roughly one lot of zero, whereas for $M=20$ lots the bands extend to around six lots. 

The algorithm therefore does not generate its cash flows by leaning to one side of the market, but by quoting on both sides while keeping its inventory centered at zero. This does not mean that it quotes symmetrically in every state, as the allocations conditional on a long position in Figure~\ref{fig:actions_imb} show.

\section{Conclusion}
We introduce a novel RL algorithm that combines deep-set encoders with a logistic-normal distribution to model limit order allocations. The algorithm observes general market states and the levels, queue positions, and sizes of resting limit orders, enabling flexible volume allocation across market and limit orders with varying price levels. It outperforms competitive benchmarks across three simulated market environments, demonstrating intelligent placement behavior and effective inventory risk management. While we work in a simulated market environment, the algorithm can be trained on any market simulator. More broadly, our method can be used for any dynamic allocation task with variable-length feature sets. 

\bibliographystyle{plainnat}
\bibliography{../references}

\appendix 

\section{Reinforcement learning algorithm details}

\subsection{Parameters}
\label{sec:rl_parameters}

\paragraph{Neural network architecture}
For the value function neural network $ f^V_\vartheta$, we use a network with two hidden layers, 128 nodes per hidden layer, tanh activations, and an output layer with one node. For the network $f^m_{\tm}$, specifying the mean of the normal distribution, we use a network with two hidden layers, 128 nodes per hidden layer, tanh activations, and an output layer with $2(K+1)$ nodes. For the encoder network $f^o_\phi$, we use a one-layer neural network with two nodes and a ReLU activation function. 

\paragraph{Initialization}
For the output layer of $f^m_{\tm},$ we set the bias to $ b = (1, 1, \dots, 1) \in \R^{2(K+1)} $ and initialize the other weights using orthogonal initialization with a gain factor $10^{-5}$; see, e.g.,~\cite{saxe2013exact}. Therefore, the weights of the output layer of $f^m_{\tm}$ are close to zero, except for the bias term, and the output is determined by the bias term. We initialize the weights of all other layers of $f^V_\vartheta$, $f^m_{\tm}$, and $f^o_\phi$ with an orthogonal initialization scheme with a gain factor of $\sqrt{2}$ and a bias term set to zero. The parameters of the covariance matrix in \eqref{eq:covariance} are initialized as $\theta^{v,1} = \theta^{v,2} = \dots = \theta^{v,2(K+1)} = 0.$

\paragraph{Learning}
To update the encoder, actor, and critic parameters, we collect a batch of $\tau=\text{1,280}$ episodes \eqref{eq:batch_of_trajectories}. We use the Adam optimizer \citep{kingma2017adam} with a learning rate of $\eta=0.0005.$ We update the parameters for a total of $H=800$ gradient steps using the loss function \eqref{eq:policy_gradient_loss} together with the parameter $c_V=0.5$. We use the inventory risk parameter $\gamma=0.01$ in \eqref{eq:rewards}. All parameters related to the algorithm are summarized in Table~\ref{table:rl_hyperparameters}.

\paragraph{Computational resources}
The experiments were conducted on a workstation equipped with 128 CPUs and a single NVIDIA GeForce RTX GPU with 24 GB VRAM. 

\begin{table}[htbp] 
	\caption{Parameters of the RL algorithm.}
	\label{table:rl_hyperparameters}
	\vskip 0.15in
	\begin{center}
		\begin{small} 	
			\begin{sc}
				\begin{tabular}{l | c }
					\toprule 
					Gradient steps $H$ & $ 800 $ \\ 						
					Number of trajectories $\tau$ & 1,280 \\
					Learning rate $\eta$ & $0.0005$ \\ 			
                    Number of hidden layers $f^V_\vartheta$ & 2 \\
                    Nodes $f^V_\vartheta$ & 128 \\ 
                    Number of hidden layers $f^m_\tm$ & 2 \\
                    Nodes $f^m_\tm$ & 128 \\ 
					 Number of hidden layers $f^o_\phi$ & 1 \\
					Nodes $f^o_\phi$ & 2 \\
                    Loss parameter $c_V$ & $0.5$ \\ 
                    Inventory parameter $\gamma$ & $0.01$ \\ 
					\bottomrule
				\end{tabular}
			\end{sc}
		\end{small}
	\end{center}
	\vskip -0.1in
\end{table}

\subsection{Feature normalization}
\label{sec:feature_normalization} 

The features described in Section~\ref{sec:state_space} are normalized. We transform the features so that they are approximately contained in the interval $[-1,1]$ or $[0,1]$. 
\paragraph{Market states}
\begin{itemize}
	\item We use the best bid and ask price returns $100\times (p^b_n-p^b_0)/p^b_0$ and $100\times (p^a_n-p^a_0)/p^a_0$. 	
	\item According to Figure~\ref{fig:shape}, the average queue sizes are always smaller than 100. Therefore, we use the normalized volumes $v^{b,k}_n/100 \in \R_+ $ and $v^{a,k}_n/100 \in \R_+ $ for $k \in \{1,2,\dots,K\}.$
	\item We normalize the market order flow $\Delta^M_n$ in $(t_{n-1}, t_n]$ by dividing it by the total volume of market buy and sell orders in $(t_{n-1}, t_n].$
	\item We normalize the limit order flow $\Delta^L_n$ in $(t_{n-1}, t_n]$ by dividing it by the total volume of limit buy and sell orders in $(t_{n-1}, t_n]$.
    \item We normalize the cancellation order flow $\Delta^C_n$ in $(t_{n-1}, t_n]$ by dividing it by the total volume of limit buy and sell order cancellations in $(t_{n-1}, t_n].$
	\item We use the mid-price returns relative to the last mid-price $100\times(p_n-p_{n-1})/p_{n-1}$. 		
\end{itemize}

\paragraph{Private states}

\begin{itemize}
	\item We normalize the current time relative to the terminal time by considering $t_n/T \in [0,1].$ 
	\item We normalize the inventory by $\inventory_n/M \in \R.$ 
    \item For $i\in\{1,2,\dots, \activeorders^b_n\}$, we use the following normalized queue positions and sizes of the resting limit buy orders
    \begin{equation}
    \left(\frac{q^{b,i}_n}{100}, \frac{w^{b,i}_n}{M}\right) \in \R_+ \times \R_+.
    \end{equation}
     We normalize $q^{b,i}_n$ by $100,$ because Figure~\ref{fig:shape} indicates that average queue sizes are less than $100.$ Finally, order sizes cannot be larger than $M,$ which justifies the normalization $\frac{w^{b,i}_n}{M}.$ 
    \item Similarly, for $i\in\{1,2,\dots, \activeorders^a_n\}$, we use the following normalized queue positions and sizes of the resting limit sell orders
    \begin{equation}
    \left(\frac{q^{a,i}_n}{100}, \frac{w^{a,i}_n}{M}\right) \in \R_+ \times \R_+.
    \end{equation}
\end{itemize}

\section{Simulation details}

\subsection{Simulation parameters}
\label{sec:simulation_parameters}
\begin{figure}[htbp]
	\centering
	\begin{subfigure}[t]{0.3\textwidth}
		\includegraphics[width=\textwidth]{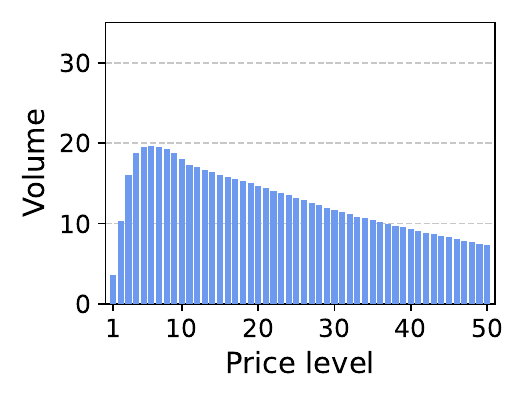}
		\caption{Noise}
		\label{fig:shape_noise}
	\end{subfigure}
	\hfill
	\begin{subfigure}[t]{0.3\textwidth}
		\includegraphics[width=\textwidth]{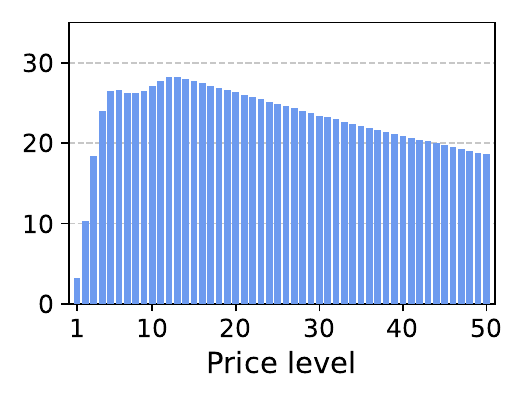}
		\caption{Noise \& Tactical}
		\label{fig:shape_tactical}
	\end{subfigure}
	\hfill 
	\begin{subfigure}[t]{0.3\textwidth}
	\includegraphics[width=\textwidth]{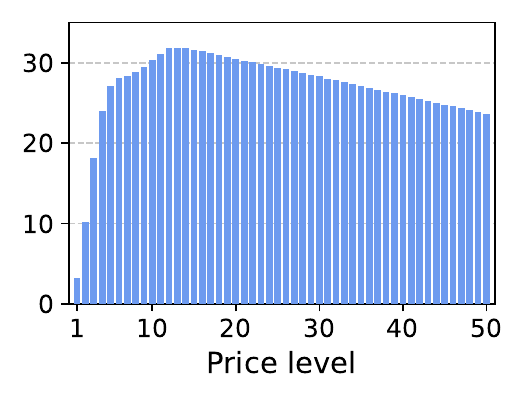}
	\caption{Noise, Tactical \& Strategic}
	\label{fig:shape_strategic}			
	\end{subfigure}	
	\caption{Average order book shapes for the three market environments.}
	\label{fig:shape}
\end{figure}

In this section, we discuss all parameters related to the trading agents. Noise, tactical, and strategic traders submit orders up to level $D=30.$ The order sizes of the trading agents have a half-normal distribution with the same standard deviations $\traderstd^M_{\text{noise}} = \traderstd^{L,k}_{\text{noise}} = \traderstd^{C,k}_{\text{noise}} =2,$ $\traderstd^M_{\text{tactical}} = \traderstd^{L,k}_{\text{tactical}} = \traderstd^{C,k}_{\text{tactical}} =2$ and $\traderstd^M_{\text{strategic}} = \traderstd^{L,k}_{\text{strategic}} = \traderstd^{C,k}_{\text{strategic}} =2$ for $k \in \{1, 2, \dots,D\}.$ The agents start trading at time $-\Delta t=-30\,\mathrm{s},$ where the best bid price is $p^b_{-\Delta t}=1000$ and the best ask price is $p^a_{-\Delta t}=1001,$ and the order book is populated with limit orders. We start each trading simulation from its corresponding average shape, which is displayed in Figure~\ref{fig:shape}. More precisely, at $-\Delta t,$ we set the volume $k-1$ ticks below or above the best prices to $\bar{v}^{k},$ for $k=1,\dots,D,$ where $\bar{v}^{k}$ is the average volume, defined in \eqref{eq:average_shape} below. All simulation parameters are summarized in Table~\ref{table:simulation_parameters}. 
\begin{table}[htbp] 
	\caption{Parameters for the market simulation.}
	\label{table:simulation_parameters}
	\begin{center}
		\begin{small} 	
			\begin{sc}
				\begin{tabular}{l | c }
					\toprule 
					start time for simulation $-\Delta t$ & $-30\,\mathrm{s}$ \\ 
                    end time for simulation $T$ & $600\,\mathrm{s}$ \\ 
                    initial bid price $p^{b}_{-\Delta t}$ & 1000 \\
                    initial ask price $p^{a}_{-\Delta t}$ & 1001 \\
                    initial volumes & Figure~\ref{fig:shape} \\
                    order size $\traderstd^M_{\text{\normalfont noise}} = \traderstd^{L,k}_{\text{\normalfont noise}} = \traderstd^{C,k}_{\text{\normalfont noise}} $ & 2 \\
                    order size 
                    $\traderstd^M_{\text{\normalfont tactical}} = \traderstd^{L,k}_{\text{\normalfont tactical}} = \traderstd^{C,k}_{\text{\normalfont tactical}} $ & 2 \\ 
                    order size 
                    $\traderstd^M_{\text{\normalfont strategic}} = \traderstd^{L,k}_{\text{\normalfont strategic}} = \traderstd^{C,k}_{\text{\normalfont strategic}} $ & 2 \\ 
					market order intensity $\lambda^M$ & $0.1237$ \\ 
					Limit/Cancel intensities $\lambda^{L,k}, \lambda^{C,k}$ & Table~\ref{table:intensities} \\						
					Tactical trader factor $d^M = d^{L,k} = d^{C,k} $ & 4 \\ 
                    Strategic trader factor $z^M = z^{L,k} = z^{C,k} $ & 2 \\ 
                    Imbalance smoothing factor $ \beta $ & 0.1 \\ 
					damping factor $c$ & 0.65 \\
					start time for RL and benchmark algorithms & $t=0$ \\ 
					small maximum lots $M$ & 2 \\
					large maximum lots $M$ & 20 \\
					\bottomrule
				\end{tabular}
			\end{sc}
		\end{small}
	\end{center}
	\vskip -0.1in
\end{table}

\paragraph{Average shapes} 
We start the market simulation from an equilibrium state of the limit order book. The equilibrium states are described by the average volumes relative to the best price
\begin{equation}
\bar{v} = (\bar{v}^{1}, \dots, \bar{v}^{D})
\in \R_+^{D}.
\label{eq:average_shape} 
\end{equation}
Here, for $k\in\{1,\dots,D\}$ the quantity $\bar{v}^{k}$ is the average volume $k-1$ ticks below or above the best prices. The average shapes are obtained by running simulations for a long time and averaging the volumes at each price level. In our simulations we always start with initial prices $p^b_{-\Delta t}=1000$ and $p^a_{-\Delta t}=1001.$ The initial volumes at the best prices are then $\bar{v}^1,$ the initial volumes at the second-best prices are $\bar{v}^2,$ and so on. The average shape for each market is summarized in Figure~\ref{fig:shape}.

\paragraph{Market with noise traders}
The intensities of limit order $\lambda^{L,k}$ and cancellation $\lambda^{C,k}$ arrivals for the noise traders are presented in Table~\ref{table:intensities}. The market order intensity is given by $\lambda^M=0.1237.$ The intensities are taken from the paper \cite{abergel2013mathematical}, but we increase the cancellation intensities by a factor of $100.$ This leads to smaller queue sizes, which makes the simulation more efficient. 
At $- \Delta t$, the order book is populated with limit buy and sell orders, such that the shape of the order book matches its average shape, shown in Figure~\ref{fig:shape_noise}.

\begin{table}[htbp]
	\caption{Intensities of limit orders and cancellations.}
	\label{table:intensities}
	\begin{center}
		\begin{small} 	
			\begin{sc}
				\begin{tabular}{c c c }
					\toprule
					$k$ (ticks) & $\lambda^{L,k}$ & $10 \lambda^{C,k}$ \\
					\midrule
					1 & 0.2842 & 0.8636 \\
					2 & 0.5255 & 0.4635 \\
					3 & 0.2971 & 0.1487 \\
					4 & 0.2307 & 0.1096 \\
					5 & 0.0826 & 0.0402 \\
					6 & 0.0682 & 0.0341 \\
					7 & 0.0631 & 0.0311 \\
					8 & 0.0481 & 0.0237 \\
					9 & 0.0462 & 0.0233 \\
					10 & 0.0321 & 0.0178 \\
					11 & 0.0178 & 0.0127 \\
					12 & 0.0015 & 0.0012 \\ 
					13 & 0.0001 & 0.0001 \\ 
					14 & 0.0000 & 0.0000 \\ 
					\vdots & \vdots & \vdots \\ 
					30 & 0.0000 & 0.0000 \\
					\bottomrule
				\end{tabular}
			\end{sc}
		\end{small}
	\end{center}	
\end{table}

\paragraph{Market with noise \& tactical traders}
The intensities of the order arrivals from the tactical traders depend on the exponentially weighted volumes defined in \eqref{eq:weighted_volumes}. We set the damping factor to $c=0.65.$ We set the parameters that control the sensitivity of the order arrivals to the weighted volume imbalance to $d^{L,k} = d^{C,k} = d^M = 4$ for $k\in\{1,2, \dots, D\}.$ To compensate for the presence of tactical traders, we reduce the intensities of the noise traders by 30\%. At the initial time $-\Delta t,$ the order book is populated with limit orders, such that its shape matches the equilibrium state, displayed in Figure~\ref{fig:shape_tactical}.

\paragraph{Market with noise \& tactical \& strategic traders}
The intensities of the order arrivals from the strategic traders depend on the smoothed imbalance signal, defined in \eqref{eq:smoothed_imbalance}. We set the smoothing parameter to $\beta=0.1$ and the sensitivity parameters, which control how strongly the strategic traders react to the smoothed imbalance signal, to $z^{L,k} = z^{C,k} = z^M = 2$ for $k\in\{1,2, \dots, D\}.$ To compensate for the additional flow from tactical and strategic traders, we reduce the intensities of the noise traders by 40\%. At the initial time $-\Delta t,$ the order book is populated with limit orders, such that its shape matches the equilibrium state, displayed in Figure~\ref{fig:shape_strategic}.

\section{Additional numerical experiments}
\label{sec:additional_experiments}

\subsection{Dirichlet vs.\ logistic-normal distribution}
\label{sec:dirichlet_vs_logistic_normal}

An alternative to the logistic-normal distribution is the Dirichlet (DR) distribution, which is also defined on the probability simplex $\SI^{2(K+1)}.$ It has also been used in \cite{cheridito2026reinforcement} for a trade-execution problem. Let us define
$\alpha = (\alpha^0, \alpha^1, \alpha^2, \dots, \alpha^{2(K+1)})\in\R^{2K+3}$ with $\alpha^k > 0,$ for $k=0,1,\dots,2(K+1),$ and $\bar{\alpha} = \sum_{k=0}^{2(K+1)} \alpha^k.$ The density of the Dirichlet distribution is given by 
\[
\psi(\action^0, \action^1, \dots, \action^{2(K+1)}) = \frac{1}{B(\alpha)} \prod_{k=0}^{2(K+1)} (\action^k)^{\alpha^k - 1}
\;
\text{ with }
\;
B(\alpha) = \frac{\prod_{k=0}^{2(K+1)} \Gamma(\alpha^k)}{\Gamma\left( \bar{\alpha} \right)}.
\]
If $\action$ is a random variable with a Dirichlet distribution, then the expected value for each component is given by 
\begin{align}
\label{eq:expected_value_dirichlet}
\E[\action^k] &= \frac{\alpha^k}{\bar{\alpha}},
\quad
\text{for } k=0,\dots,2(K+1).
\end{align}

Instead of using the logistic-normal distribution to parameterize the policy, we can use the Dirichlet distribution. The resulting algorithm is trained with the same hyperparameters as the LN algorithm. We also use the same encoder architecture and the same value function architecture as in the LN algorithm. The policy network again has two hidden layers, 128 nodes per hidden layer, and tanh activations. The only difference is the output layer: the network (which we called $f^m_{\tm}$ above) has $2K+3$ nodes, corresponding to the parameters $\alpha^0, \alpha^1, \dots, \alpha^{2(K+1)}$ of the Dirichlet distribution. To ensure that the parameters are positive, we use the softplus function as an activation function in the output layer. We use an orthogonal initialization scheme using a gain factor of $10^{-5},$ leading to small weights. We set the bias of the final layer to $b=\operatorname{softplus}^{-1}(1,10, \dots, 10, 10)\in\R^{2K+3},$ where the inverse is applied to each component. Therefore, after applying the softplus activation, we have
\[
\E[\action^k] \approx 10\,\E[\action^0] \quad \text{for } k = 1,2,\dots,2(K+1).
\]
This means that the actions $\action^k$ for $k=1,2,\dots,2(K+1),$ which correspond to placing market or limit orders, are initially more likely than action $\action^0,$ which corresponds to not placing any order.

The results of the experiment are reported in Table~\ref{table:pnl_ln_vs_dr}. We use the same evaluation procedure as described in Section~\ref{sec:evaluation_and_rl_setup}. In particular, we use 10,000 test episodes per environment and per value of $M$ to compute expected values and standard deviations of the normalized cash flow \eqref{eq:sum_of_rewards_normalized}. We use the trained policy network to produce parameters $(\alpha^0, \alpha^1, \dots, \alpha^{2(K+1)})$ at each state, and sample actions from the resulting Dirichlet distribution. Across both values $M=2$ and $M=20,$ the LN algorithm has higher expected cash flow than the DR algorithm in all three market environments. For the market with noise traders, the LN algorithm also has higher standard deviation than the DR algorithm. For the other two markets, the LN algorithm has notably lower standard deviation than DR for $M=20$, while for $M=2$ the standard deviations are close in the market with noise and tactical traders and higher for LN in the market with noise, tactical, and strategic traders. The gap in expected cash flow is large in the latter environment. Overall, we see strong outperformance of the LN algorithm compared to the DR algorithm, especially in the more complex market environments with tactical and strategic traders.
\begin{table}[htpb]
\begin{center}
    \begin{scriptsize}
        \begin{sc}
\begin{tabular}{llcccc}
\toprule
 & Lots & $\E[\text{DR}]$ & $\sigma[\text{DR}]$ & $\E[\text{LN}]$ & $\sigma[\text{LN}]$ \\
\midrule
noise & 2 & 4.87 & 2.20 & \textbf{6.03} & 2.81 \\
 & 20 & 3.99 & 1.33 & \textbf{4.66} & 1.41 \\
noise \& tactical & 2 & 6.72 & 2.27 & \textbf{9.11} & 2.20 \\
 & 20 & 1.63 & 1.88 & \textbf{5.68} & 1.03 \\
noise \& tactical & 2 & 4.87 & 1.77 & \textbf{8.55} & 2.40 \\
\& strategic & 20 & $-6.18$ & 1.60 & \textbf{4.99} & 1.07 \\
\bottomrule
\end{tabular}
        \end{sc}
    \end{scriptsize}
\end{center}
\caption{Expected value and standard deviation of the normalized cash flow \eqref{eq:sum_of_rewards_normalized} under the Dirichlet (DR) and logistic-normal (LN) policies, across three market environments and for the two cases $M=2$ and $M=20$ lots. The highest expected value in each row is highlighted in bold.}
\label{table:pnl_ln_vs_dr}
\end{table}

\begin{figure}[htbp]
	\centering
	\begin{subfigure}[t]{0.3\textwidth}
		\includegraphics[width=\textwidth]{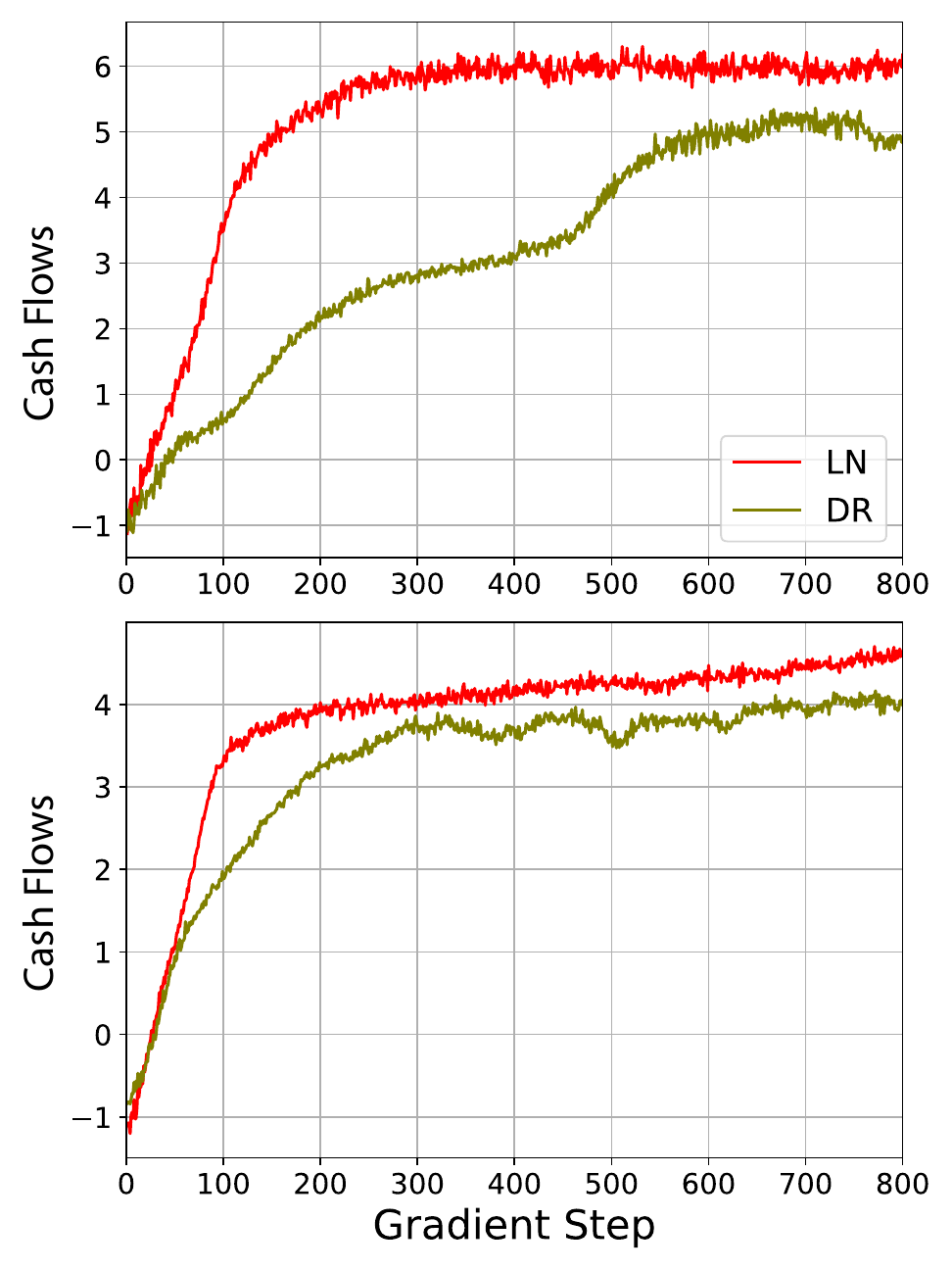}
		\caption{Noise}
	\end{subfigure}
	\hfill
	\begin{subfigure}[t]{0.3\textwidth}
		\includegraphics[width=\textwidth]{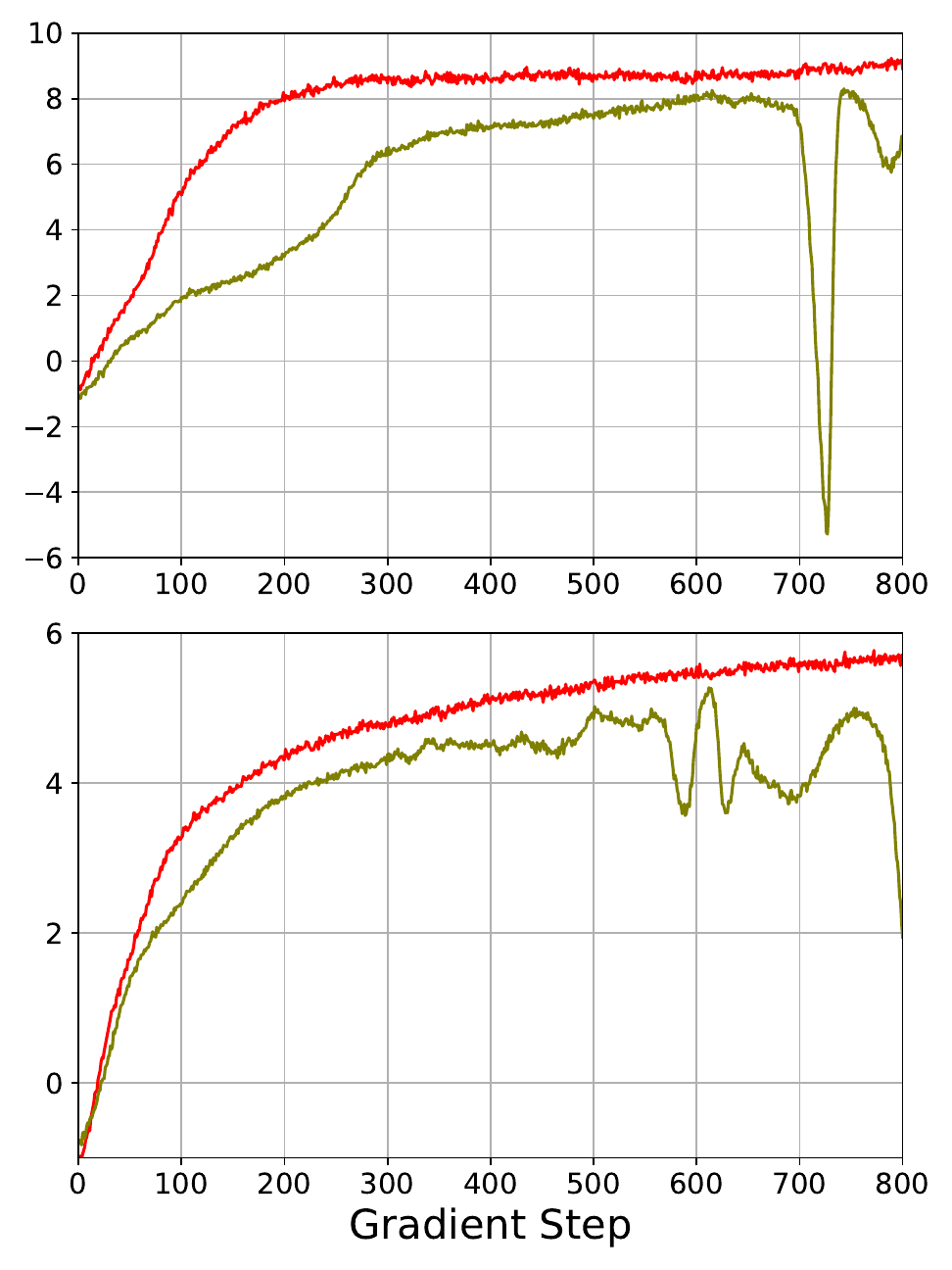}
		\caption{Noise \& Tactical}
	\end{subfigure}
	\hfill
	\begin{subfigure}[t]{0.3\textwidth}
		\includegraphics[width=\textwidth]{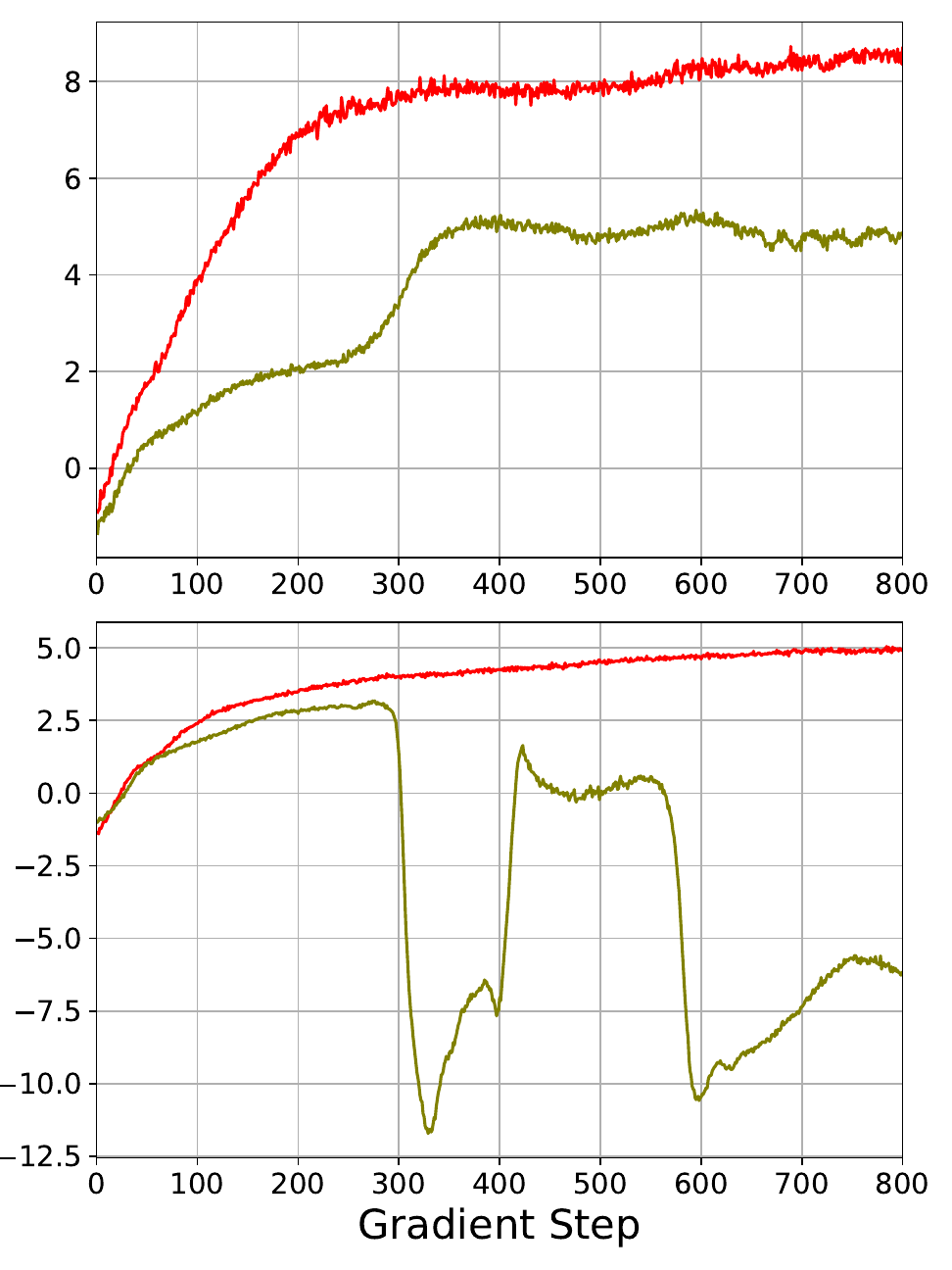}
		\caption{Noise, Tactical \& Strategic}
	\end{subfigure}
	\caption{Average normalized cash flow \eqref{eq:sum_of_rewards_normalized} per gradient step for the LN and DR algorithms across three market environments. The upper panels correspond to $M=2$ lots, while the lower panels correspond to $M=20$ lots. 
	}
	\label{fig:return_convergence_ln_vs_dr}
\end{figure}

Figure~\ref{fig:return_convergence_ln_vs_dr} shows the average normalized cash flow \eqref{eq:sum_of_rewards_normalized} per gradient step for the LN and DR algorithms across three market environments, where the average is computed across the batch of trajectories used for one gradient step. The upper panels correspond to $M=2$ lots, while the lower panels correspond to $M=20$ lots. We see that the LN algorithm converges faster and to a higher value than the DR algorithm across all three market environments and both values of $M$. Furthermore, the DR algorithm suffers from instabilities that are especially pronounced in the more complex market environments with tactical and strategic traders. In contrast, the LN algorithm is stable in training across all three market environments.

\subsection{Effect of inventory penalty}
\label{sec:effect_of_inventory_penalty}

\begin{figure}[htbp]
	\centering
	\begin{subfigure}[t]{0.3\textwidth}
		\includegraphics[width=\textwidth]{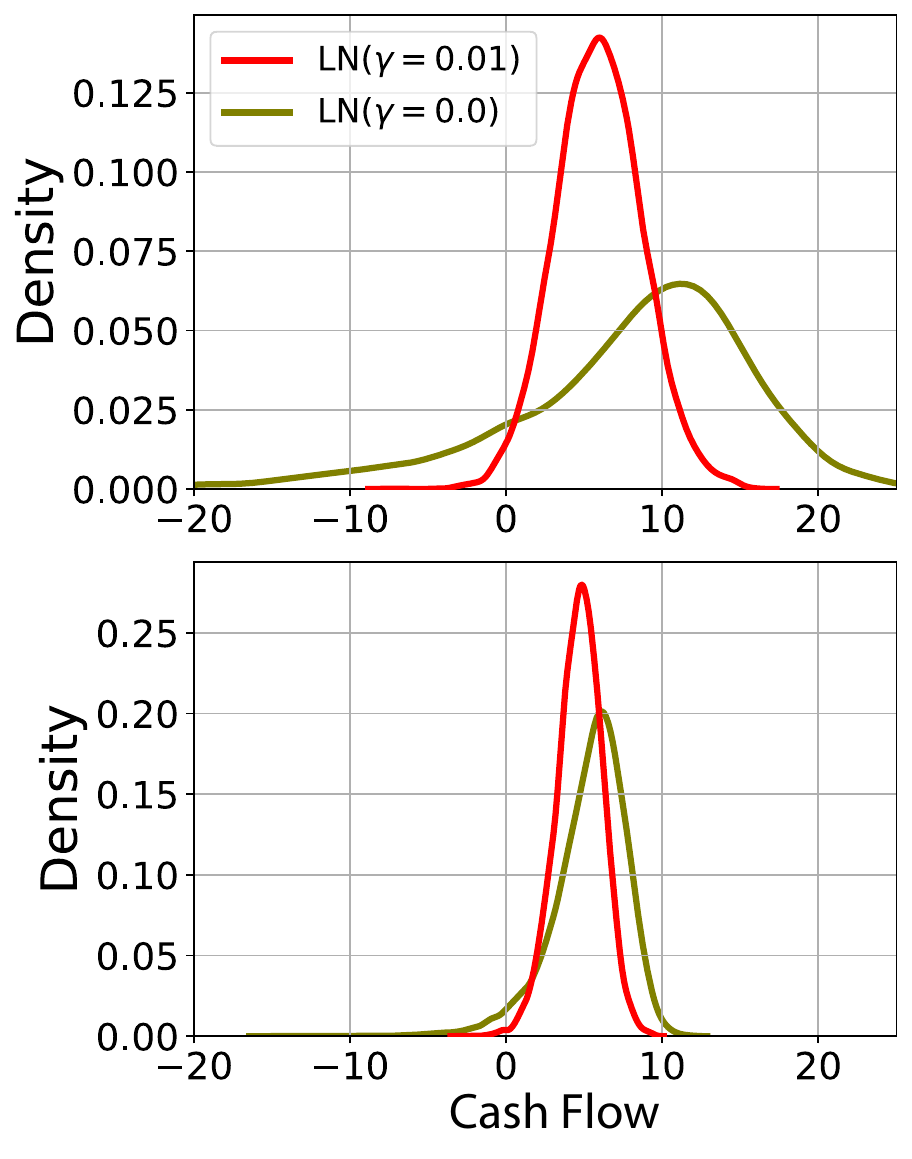}
		\caption{Noise}
	\end{subfigure}
	\hfill
	\begin{subfigure}[t]{0.3\textwidth}
		\includegraphics[width=\textwidth]{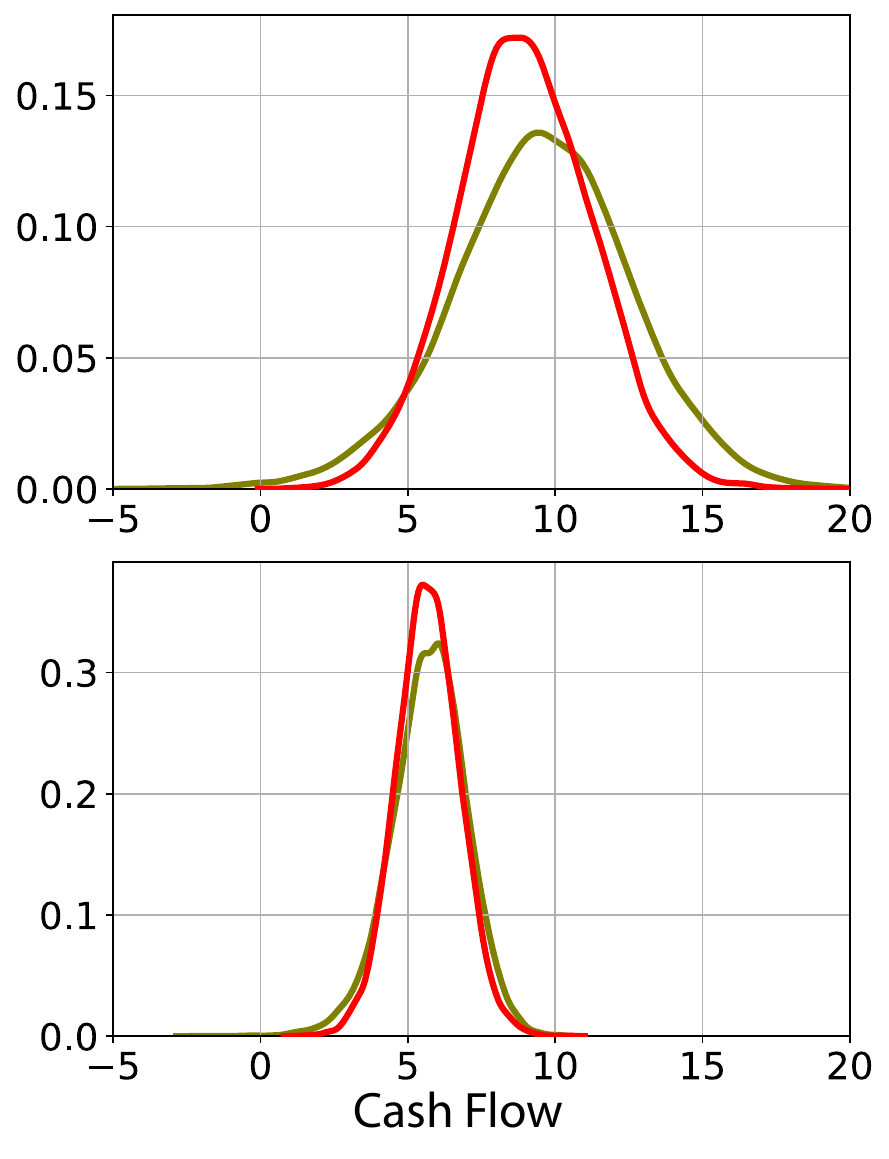}
		\caption{Noise \& Tactical}
	\end{subfigure}
	\hfill
	\begin{subfigure}[t]{0.3\textwidth}
		\includegraphics[width=\textwidth]{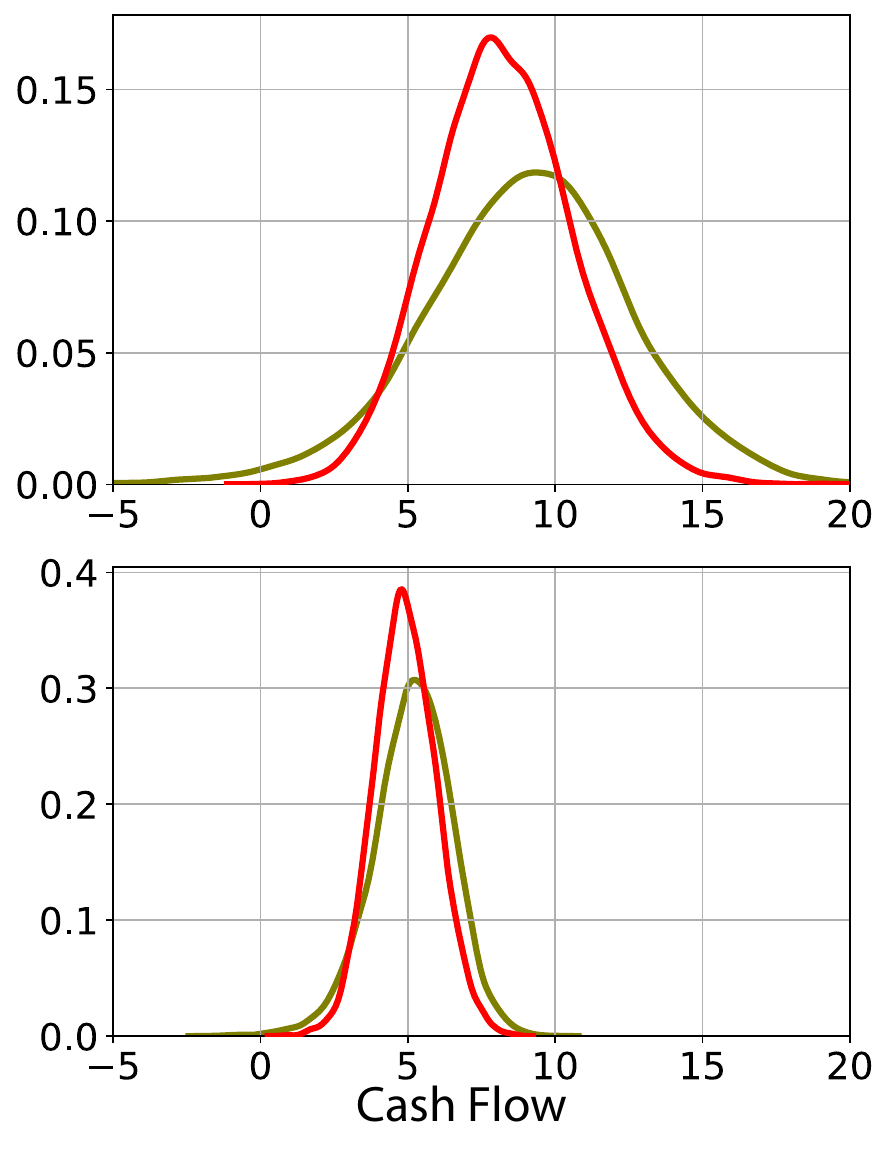}
		\caption{Noise, Tactical \& Strategic}
	\end{subfigure}
	\caption{Histograms of normalized cash flows \eqref{eq:sum_of_rewards_normalized} for the LN algorithm with and without inventory penalty. The upper panels correspond to $M=2$ lots, while the lower panels correspond to $M=20$ lots.
	}
	\label{fig:pnl_histogram_inventory_penalty}
\end{figure}

In this section, we analyze the effect of the inventory penalty parameter $\gamma$, used in \eqref{eq:rewards}, on the performance of the LN algorithm. We compare the performance of the LN algorithm with inventory penalty ($\gamma=0.01$) and without ($\gamma=0.0$). The results are reported in Table~\ref{table:pnl_inventory_penalty} and Figure~\ref{fig:pnl_histogram_inventory_penalty}, where the two algorithms are denoted by $\text{LN}(\gamma=0.01)$ and $\text{LN}(\gamma=0.0)$. Note that in Section~\ref{sec:numerical_experiments}, we denoted the LN algorithm with inventory penalty $\gamma=0.01$ simply as $\text{LN}$, thereby suppressing the dependence on $\gamma$. We see that across all three market environments and both values of $M$, the $\text{LN}(\gamma=0.0)$ algorithm has higher expected cash flows together with higher standard deviations than the $\text{LN}(\gamma=0.01)$ algorithm.

Interestingly, the inventory penalty does not have the same effect across all market environments. The difference in standard deviations for the $\text{LN}(\gamma=0.01)$ and $\text{LN}(\gamma=0.0)$ algorithms is more pronounced in the market consisting solely of noise traders than in the other two markets. This effect is most striking at low inventory capacity ($M=2$), where the standard deviation drops from $8.20$ to $2.81$ once the penalty is applied, whereas for $M=20$ the reduction is comparatively modest. This suggests that in a market in which trade intensities are state-dependent, a static inventory penalty is not effective in controlling inventory. An interesting research direction is to explore a state-dependent inventory risk parameter that can adapt to market states.

\begin{table}[htpb]
\begin{center}
    \begin{scriptsize}
        \begin{sc}
\begin{tabular}{llcccc}
\toprule
 & Lots & $\E[\text{LN}(\gamma=0.01)]$ & $\sigma[\text{LN}(\gamma=0.01)]$ & $\E[\text{LN}(\gamma=0.0)]$ & $\sigma[\text{LN}(\gamma=0.0)]$ \\
\midrule
noise & 2 & 6.03 & 2.81 & \textbf{7.95} & 8.20 \\
 & 20 & 4.66 & 1.41 & \textbf{5.35} & 2.39 \\
noise \& tactical & 2 & 9.11 & 2.20 & \textbf{9.55} & 3.05 \\
 & 20 & 5.68 & 1.03 & \textbf{5.72} & 1.28 \\
noise \& tactical & 2 & 8.55 & 2.40 & \textbf{9.05} & 3.55 \\
\& strategic & 20 & 4.99 & 1.07 & \textbf{5.16} & 1.35 \\
\bottomrule
\end{tabular}
        \end{sc}
    \end{scriptsize}
\end{center}
\caption{Expected value and standard deviation of the normalized cash flow \eqref{eq:sum_of_rewards_normalized} under the logistic-normal policy with and without inventory penalty parameter, across three market environments and for the two cases $M=2$ and $M=20$ lots. The highest expected value in each row is highlighted in bold.}
\label{table:pnl_inventory_penalty}
\end{table}

\subsection{Effect of terminal inventory constraint}
\label{sec:effect_of_terminal_inventory_constraint}

In this section, we analyze the effect of the terminal inventory parameter $\nu$, which was defined in \eqref{eq:terminal_reward}. We compare two cases. The first case is $\nu=0.0$, which requires the algorithm to have zero inventory at terminal time. Any non-zero inventory at terminal time must be bought back or sold with a market order, such that $\inventory_{N+}=0.$ The second case is $\nu=0.5$, which means that the algorithm is allowed to have a terminal inventory of absolute size up to $\ceil{0.5 M}.$ Any inventory that is larger in absolute value than the limit must be bought back or sold with a market order, such that $|\inventory_{N+}| \leq \ceil{0.5 M}.$ Furthermore, the remaining inventory, of absolute size at most $\ceil{0.5 M}$, is valued at the mid-price at time $t_N.$ This mid-price valuation yields a higher terminal value than liquidating the inventory with a market order, because when using a market order, the algorithm must pay the spread and additional costs from walking the limit order book. We set the running inventory penalty defined in \eqref{eq:rewards} to $\gamma=0.0$, in order to isolate the effect of the terminal inventory parameter $\nu$.

\begin{figure}[htbp]
	\centering
	\begin{subfigure}[t]{0.3\textwidth}
		\includegraphics[width=\textwidth]{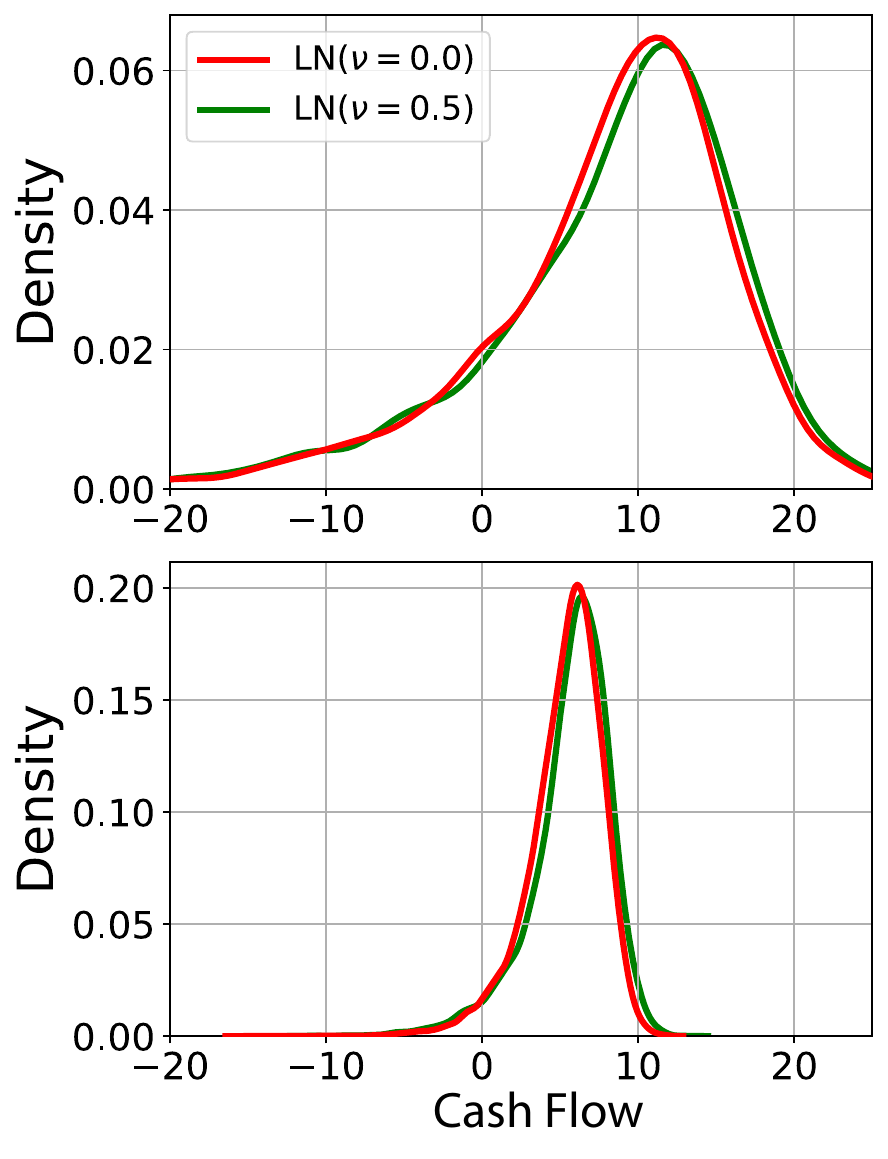}
		\caption{Noise}
	\end{subfigure}
	\hfill
	\begin{subfigure}[t]{0.3\textwidth}
		\includegraphics[width=\textwidth]{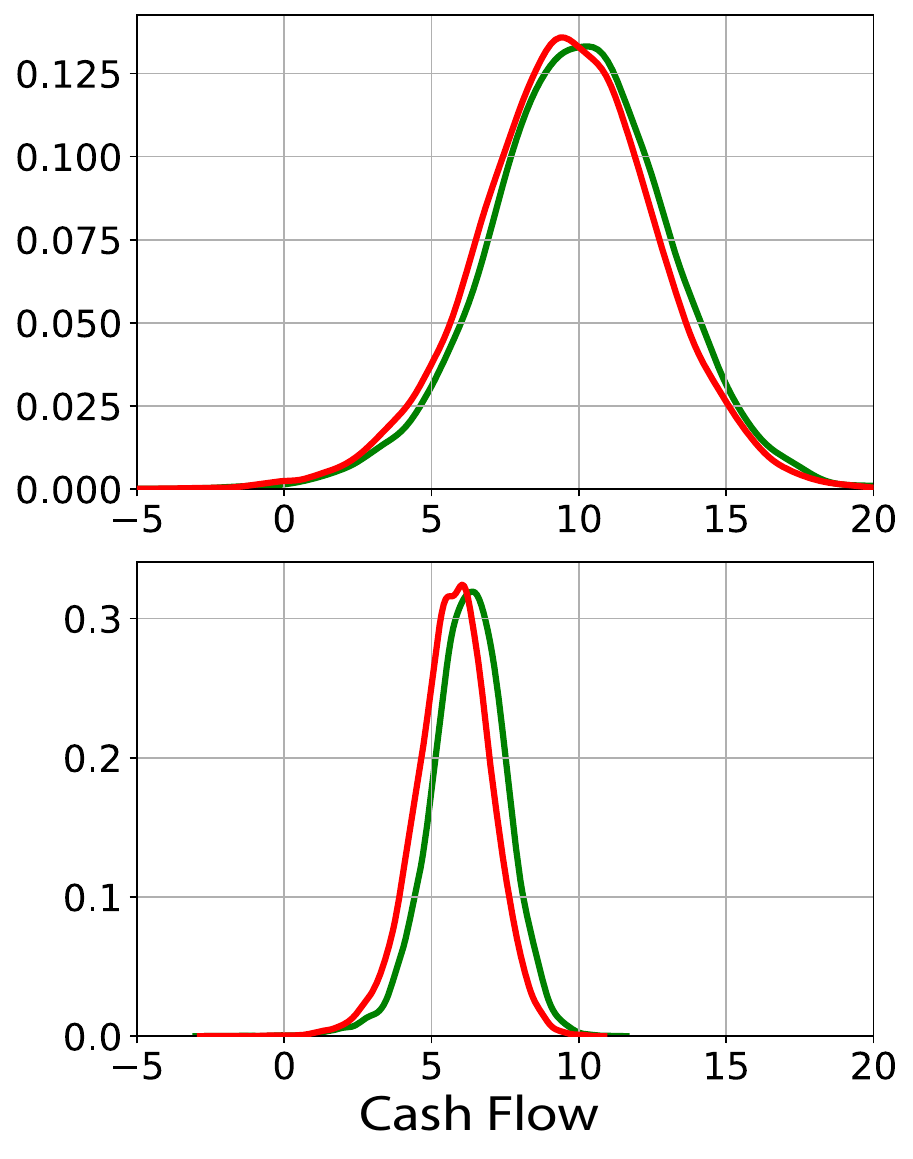}
		\caption{Noise \& Tactical}
	\end{subfigure}
	\hfill
	\begin{subfigure}[t]{0.3\textwidth}
		\includegraphics[width=\textwidth]{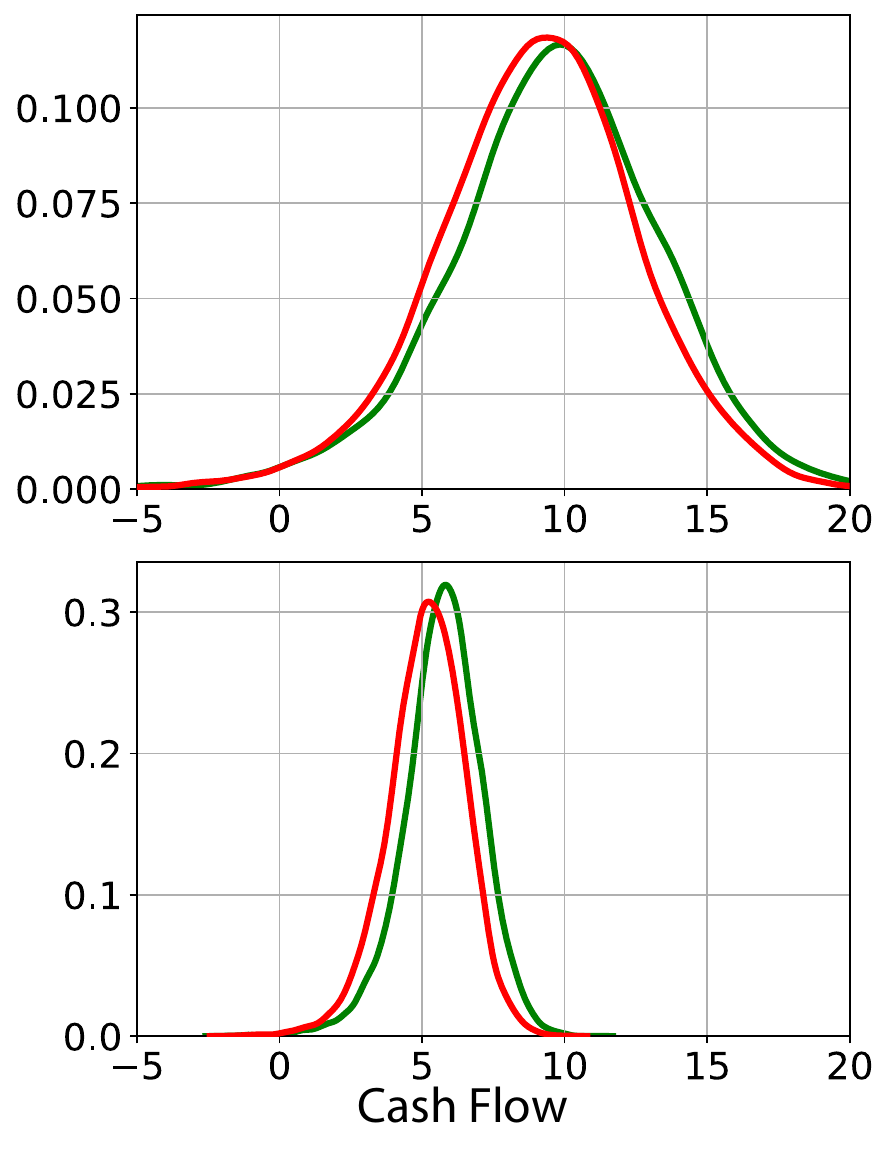}
		\caption{Noise, Tactical \& Strategic}
	\end{subfigure}
	\caption{Histograms of normalized cash flow \eqref{eq:sum_of_rewards_normalized} for the LN algorithm with running inventory penalty $\gamma=0.0$ and terminal inventory parameter $\nu=0.0$ and $\nu=0.5$. The upper panels correspond to $M=2$ lots, while the lower panels correspond to $M=20$ lots.
	}
	\label{fig:pnl_histogram_terminal_inventory_penalty}
\end{figure}

The results are reported in Table~\ref{table:pnl_terminal_penalty} and Figure~\ref{fig:pnl_histogram_terminal_inventory_penalty}, where the two algorithms are denoted by $\text{LN}(\nu=0.0)$ and $\text{LN}(\nu=0.5)$.
 Technically, we should use the notation $\text{LN}(\nu=0.0, \gamma=0.0)$ and $\text{LN}(\nu=0.5, \gamma=0.0)$, but we suppress the dependence on $\gamma$ for brevity.
We see from Table~\ref{table:pnl_terminal_penalty} that the $\text{LN}(\nu=0.5)$ algorithm has higher expected cash flows than the $\text{LN}(\nu=0.0)$ algorithm, while the standard deviations are comparable, though slightly higher for the $\text{LN}(\nu=0.5)$ algorithm. This confirms the effect described above: by retaining part of the terminal inventory at the mid-price instead of forcing a full liquidation with spread-paying market orders, the $\text{LN}(\nu=0.5)$ algorithm avoids the associated transaction costs. Since the algorithm is not forced to flatten its inventory at terminal time, it carries a larger terminal position, whose mark-to-mid valuation is exposed to price fluctuations, explaining the slightly higher standard deviations. In line with the numerical values in the table, we see in Figure~\ref{fig:pnl_histogram_terminal_inventory_penalty} that the cash flow distribution for the $\text{LN}(\nu=0.5)$ algorithm is shifted to the right compared to the $\text{LN}(\nu=0.0)$ algorithm.

\begin{table}[htpb]
\begin{center}
    \begin{scriptsize}
        \begin{sc}
\begin{tabular}{lccccc}
\toprule
 & Lots & $\E[\text{LN}(\nu=0.0)]$ & $\sigma[\text{LN}(\nu=0.0)]$ & $\E[\text{LN}(\nu=0.5)]$ & $\sigma[\text{LN}(\nu=0.5)]$ \\
\midrule
noise & 2 & 7.95 & 8.20 & \textbf{8.36} & 8.36 \\
 & 20 & 5.35 & 2.39 & \textbf{5.67} & 2.59 \\
noise \& tactical & 2 & 9.55 & 3.05 & \textbf{9.94} & 3.06 \\
 & 20 & 5.72 & 1.28 & \textbf{6.23} & 1.29 \\
noise \& tactical & 2 & 9.05 & 3.55 & \textbf{9.65} & 3.73 \\
\& strategic & 20 & 5.16 & 1.35 & \textbf{5.73} & 1.37 \\
\bottomrule
\end{tabular}
        \end{sc}
    \end{scriptsize}
\end{center}
\caption{Expected value and standard deviation of the normalized cash flow \eqref{eq:sum_of_rewards_normalized} under the logistic-normal policy with running inventory penalty $\gamma=0.0$ and terminal inventory parameter $\nu=0.0$ and $\nu=0.5,$ across three market environments and for the two cases $M=2$ and $M=20$ lots. The highest expected value in each row is highlighted in bold.}
\label{table:pnl_terminal_penalty}
\end{table}

\end{document}